\documentclass[fleqn,usenatbib]{mnras}
\usepackage[T1]{fontenc}
\usepackage{ae,aecompl}
\usepackage{graphicx}
\usepackage{natbib}
\pdfoutput=1
\usepackage{amsmath}
\usepackage{amssymb}
\usepackage{threeparttable}
\usepackage{float}
\usepackage{epstopdf}
\usepackage{newtxtext,newtxmath}
\usepackage{titlesec}

\let\oldAA\AA
\renewcommand{\AA}{\text{\normalfont\oldAA}}

\newcommand{\HeIIL}{\hbox{{\rm He}\kern 0.1em{\sc ii}\kern 0.1em{$\lambda1640$}}}
\newcommand{\HeII}{\hbox{{\rm He}\kern 0.1em{\sc ii}}}
\newcommand{\HeIIl}{\hbox{{\rm He}\kern 0.1em{\sc ii}\kern 0.1em{$\lambda1640$}}}
\newcommand{\CIIIL}{\hbox{{\rm C}\kern 0.1em{\sc iii}]\kern 0.1em{$\lambda1907,\lambda1909$}}}
\newcommand{\CIVL}{\hbox{{\rm C}\kern 0.1em{\sc iv}\kern 0.1em{$\lambda1548,\lambda1550$}}}
\newcommand{\CIII}{\hbox{{\rm C}\kern 0.1em{\sc iii}]}}
\newcommand{\MgII}{\hbox{{\rm Mg}\kern 0.1em{\sc ii}}}
\newcommand{\NeIII}{[\hbox{{\rm Ne}\kern 0.1em{\sc iii}}]}
\newcommand{\OIII}{[\hbox{{\rm O}\kern 0.1em{\sc iii}}]}
\newcommand{\OII}{[\hbox{{\rm O}\kern 0.1em{\sc ii}}]}
\newcommand{\CIV}{\hbox{{\rm C}\kern 0.1em{\sc iv}}}
\newcommand{\CII}{[\hbox{{\rm C}\kern 0.1em{\sc ii}}]}

\title[Excess HI absorption around $z\approx4$ galaxies]{Large-scale excess HI absorption around $z\approx4$ galaxies detected in a background galaxy spectrum in the MUSE eXtremely Deep Field} 

\author [J.~Matthee et al.]{
\parbox[t]{\textwidth}{
Jorryt Matthee$^{1,2}$\thanks{E-mail: mattheej@phys.ethz.ch}\thanks{Zwicky Fellow}, Christopher Golling$^{1}$, Ruari Mackenzie$^{1}$, Gabriele Pezzulli$^{3}$, \\ Simon Lilly$^{1}$, Joop Schaye$^{4}$, Roland Bacon$^{5}$, Haruka Kusakabe$^{6}$, Tanya Urrutia$^{7}$, Leindert Boogaard$^{8}$, Jarle Brinchmann$^{9, 4}$, Michael V. Maseda$^{10}$, Thibault Garel$^{6}$, Nicolas F. Bouch\'e$^{5}$ and Lutz Wisotzki$^{7}$}
\vspace*{8pt}\\
$^{1}$Department of Physics, ETH Z\"urich, Wolfgang-Pauli-Strasse 27, 8093 Z\"urich, Switzerland\\
$^{2}$Institute of Science and Technology Austria (ISTA), Am Campus 1, 3400 Klosterneuburg, Austria\\
$^{3}$ Kapteyn Astronomical Institute, University of Groningen, Landleven 12, 9747 AD Groningen, The Netherlands\\
$^{4}$  Leiden Observatory, Leiden University, PO Box 9513, 2300 RA Leiden, The Netherlands\\
$^{5}$ Univ Lyon, Univ Lyon1, Ens de Lyon, CNRS, Centre de Recherche
Astrophysique de Lyon UMR5574, F-69230, Saint-Genis-Laval,
France\\
$^{6}$ Observatoire de Genève, Université de Genève, 51 Chemin de Pégase, 1290 Versoix, Switzerland\\
$^{7}$ Leibniz-Institut für Astrophysik Potsdam (AIP), An der Sternwarte 16, 14482 Potsdam, Germany\\
$^{8}$ Max Planck Institute for Astronomy, K\"onigstuhl 17, 69117, Heidelberg, Germany \\
$^{9}$ Instituto de Astrof\'isica e Ci\^encias do Espa\c{c}o, Universidade do Porto, CAUP, Rua das Estrelas, PT4150-762 Porto, Portugal\\
$^{10}$ Department of Astronomy, University of Wisconsin-Madison, 475 N. Charter St., Madison, WI 53706, USA
}

\pubyear{2023}

\begin{document}
\label{firstpage}
\pagerange{\pageref{firstpage}--\pageref{lastpage}}
\maketitle

\begin{abstract}
Observationally mapping the relation between galaxies and the intergalactic medium (IGM) is of key interest for studies of cosmic reionization. Diffuse hydrogen gas has typically been observed in HI Lyman-$\alpha$ (Ly$\alpha$) absorption in the spectra of bright background quasars. However, it is important to extend these measurements to background galaxies as quasars become increasingly rare at high redshift and rarely probe closely separated sight-lines. Here we use deep integral field spectroscopy in the MUSE eXtremely Deep Field to demonstrate the measurement of the Ly$\alpha$ transmission at $z\approx4$ in absorption to a background galaxy at $z=4.77$. The HI transmission is consistent with independent quasar sight-lines at similar redshifts. Exploiting the high number of spectroscopic redshifts of faint galaxies (500 between $z=4.0-4.7$ within a radius of 8 arcmin) that are tracers of the density field, we show that Ly$\alpha$ transmission is inversely correlated with galaxy density, i.e. transparent regions in the Ly$\alpha$ forest mark under-dense regions at $z\approx4$. Due to large-scale clustering, galaxies are surrounded by excess HI absorption over the cosmic mean out to 4 cMpc/h$_{70}$. We also find that redshifts from the peak of the Ly$\alpha$  line are typically offset from the systemic redshift by $+170$ km s$^{-1}$. This work extends results from $z\approx2-3$ to higher redshifts and demonstrates the power of deep integral field spectroscopy to simultaneously measure the ionization structure of the IGM and the large-scale density field in the early Universe.
\end{abstract}




\section{INTRODUCTION}
Galaxies form in the peaks of the large scale density distribution of matter in the Universe. The visible light from stars and ionised gas in galaxies however only constitutes a fraction of the baryons in such over-densities. The majority of baryonic gas is diffuse and resides in the inter- and circumgalactic medium \citep[IGM, CGM; e.g.][]{CenOstriker06,vdVoort12,McQuinn16,Tumlinson17} that is observable through Lyman-$\alpha$ (Ly$\alpha$, $\lambda 1215.67 \AA$) absorption in the spectra of bright distant (background) sources such as quasars \citep[e.g.][]{GP65,Steidel10}.

It is of key interest to observationally map the interplay between variations in the Ly$\alpha$ transmission and the distance to galaxies. The Ly$\alpha$ transmission is modulated both by density effects and ionization effects. A higher neutral gas density leads to a lower transmission, whereas a higher ionization fraction increases the transmission. The cross-correlation between galaxies and the Ly$\alpha$ transmission is therefore sensitive to the impact of feedback from star and supermassive black hole formation and growth on intergalactic gas \citep[e.g.][]{Theuns2002,Kollmeier2003,Viel13,Nagamine21} and local enhancements in the ionization field around galaxies and quasars \citep[e.g.][]{Worseck2006,Meyer20,Christenson21,Ishimoto22,Kashino22,Kakiichi23}.  

Various observational campaigns have focused on observing the redshift $z\approx2-3$ window, where quasars are abundant, the Ly$\alpha$ forest can efficiently be observed with ground-based telescopes and galaxies can be identified with the well-known Lyman-break \citep[e.g.][]{Adelberger05,Crighton11,Hayashino19}. Excess HI absorption has been detected out to several megaparsecs from typical L$^{\star}$ galaxies and quasars \citep{Steidel10,Rakic12,Rudie12,Prochaska13,Tummuangpak14,Mukae20,Liang21,Muzahid21,Horowitz22,Lofthouse23} indicating that density effects dominate the cross-correlation signal between galaxy distance and Ly$\alpha$ transmission at $z\approx2-3$. Consequentially, the presence of strong excess HI absorption has also been used to identify over-densities of hypothetical dusty galaxies that are challenging to find with typical Lyman-break or Ly$\alpha$ selection methods \citep{Newman22}.

However, due to local variations in the cosmic ionizing background at the end stages of cosmic reionization that are spatially correlated with galaxies, it is expected that the cross-correlation signal between galaxies and the Ly$\alpha$ transmission may show an opposite sign at a specific distance scale during or just after the epoch of reionization \citep[e.g.][]{Kakiichi18,Garaldi22}. This effect is somewhat similar to the stronger proximity effect that has been seen around quasars \citep[e.g.][]{Bajtlik88,Schirber04,Goncalves08}, but its excess transmission over the cosmic mean is likely much lower. Based on {\it JWST} data in a single quasar sight-line, \cite{Kashino22} recently reported a detection of excess transmission at high-redshift $z\approx6$ on distances $\sim5$ cMpc/h from galaxies with UV luminosities M$_{\rm UV}\approx-19$. Whether the excess ionization is due to the detected galaxies, or undetected galaxies that are clustered around the brighter ones, is unclear. 

Measuring the cross-correlation signal between the density field and the IGM transmission at intermediate redshifts $z\approx3-5$ could help to better differentiate density effects from (excess) ionization effects. This could in turn help to better characterise the timing of reionization and the properties of the ionizing sources by improving our interpretation of $z\sim6$ measurements. Low mass galaxies are the least biased tracers of the density field and at $z\approx3-5$ most effectively identified with deep Integral Field Spectroscopy of their Ly$\alpha$ line, for example with the wide-field Multi Unit Spectroscopic Explorer (MUSE; \citealt{Bacon10}) on the Very Large Telescope. Recent MUSE observations targeting bright $z\approx3$ quasars have been very efficient in picking up on the order of 100 faint galaxies through their Ly$\alpha$ emission \citep[e.g.][]{Mackenzie2019,Muzahid21,Lofthouse23}, but very deep MUSE observations of quasars at $z\approx5$ do not yet exist.

In this paper, we use data from the blank MUSE eXtremely Deep Field (MXDF; \citealt{Bacon21,Bacon22}) to 
measure the Ly$\alpha$ transmission in the spectrum of a background galaxy at $z\approx5$ (magnitude $\approx25.5$) instead, and cross-correlate the transmission with the projected and line of sight distance to foreground galaxies at $z\approx4$. The MXDF is the deepest IFU observation to date (140 hrs of exposure time; 3$\sigma$ point source sensitivity of $10^{-19}$ erg s$^{-1}$ cm$^{-2}$) and offers a glimpse of the capabilities of future 40m-class telescopes. The data are particularly suited to obtain sensitive continuum spectra of bright galaxies with intermediate resolution ($R\approx3000$) and simultaneously for identifying galaxies down to UV luminosities M$_{\rm UV}\approx-15$ \citep{Maseda18}. The MXDF field is roughly located in the middle of a larger mosaic of MUSE pointings with a wedding-cake layered exposure time ranging from 1 to 30 hrs \citep{Bacon17}, which further provides foreground galaxies at somewhat larger impact parameters.

The background galaxy, ID53 at $z=4.77$ \citep{Matthee22}, is the continuum-brightest galaxy at $z>3$ in the MXDF while it has a typical L$^{\star}_{\rm UV}$ luminosity. The deep spectrum and the available infrared photometry allowed detailed modeling of the young stellar population and dust attenuation, yielding a good fit with the first estimate of the stellar metallicity in such a high-redshift galaxy \citep{Matthee22}. The accurate SED fit simultaneously provides the intrinsic spectrum in the $\lambda_0 =1026-1216$ {\AA} region without attenuation through the intergalactic medium \citep[e.g.][]{Inoue14} allowing us to measure the Ly$\alpha$ transmission at $z\approx3.9-4.7$. Thus, the combination of this spectrum and the very large number of known spectroscopic redshifts of UV faint galaxies in the foreground at $z\approx4$, yields the opportunity to extend Ly$\alpha$ forest measurements using galaxies and foreground galaxy cross-correlation studies to higher redshifts, paving the way for more extensive studies in the future.

The paper is structured as follows. In Section $\ref{sec:methods}$ we describe how we measured the Ly$\alpha$ transmission from the background spectrum (\S $\ref{sec:backgroundgal}$), the foreground galaxy sample (\S $\ref{sec:foreground}$) and we describe how we reconstructed the 3D galaxy density field using the foreground sample (\S $\ref{sec:densmethod}$). In Section $\ref{sec:tauevo}$ we present our measured HI optical depth and compare it to independent measurements based on quasar sight-lines. We investigate the relation between the HI transmission and the density along our sight-line in Section $\ref{sec:denstrans}$. In Section $\ref{sec:CC}$ we present the cross-correlation between HI transmission and the distance to galaxies. We summarise our results in Section $\ref{sec:summary}$.

Throughout the paper we use a flat $\Lambda$DM cosmology with H$_0 = 70$ km s$^{-1}$ Mpc$^{-1}$ and $\Omega_M=0.3$. Magnitudes are in the AB system. Transverse distances are typically written in cMpc/h$_{70}$ where h$_{70}=0.7$. At redshift $z=4.3$, the average redshift of our sample, an impact parameter of 40 arcsec (approximately distance between the edge of the MXDF and the background galaxy) corresponds to 270 pkpc (1 cMpc/h$_{70}$), while the spectral resolution of $R\approx3000$ corresponds to 100 km s$^{-1}$, or a line-of-sight distance of $\sim1$ cMpc/h$_{70}$ under the Hubble flow.

\section{METHODS} \label{sec:methods}
The analysis in this paper is based on the spectrum of a relatively bright background galaxy ($\approx25.5$ AB magnitude) at $z=4.774$ \citep{Matthee22}, which we use to measure the HI Ly$\alpha$ optical depth at $z=3.95-4.72$, and a catalog of 504 spectroscopically confirmed foreground galaxies in this redshift range \citep{Urrutia19,Bacon22}. Here we detail the spectrum of the background galaxy and the intrinsic (i.e. pre-IGM) model that we use to measure the opacity, followed by the presentation of the properties of the foreground galaxy sample.

\subsection{Transmission measurement in the background galaxy} \label{sec:backgroundgal}

\begin{figure}
    \centering
    \includegraphics[width=8.5cm]{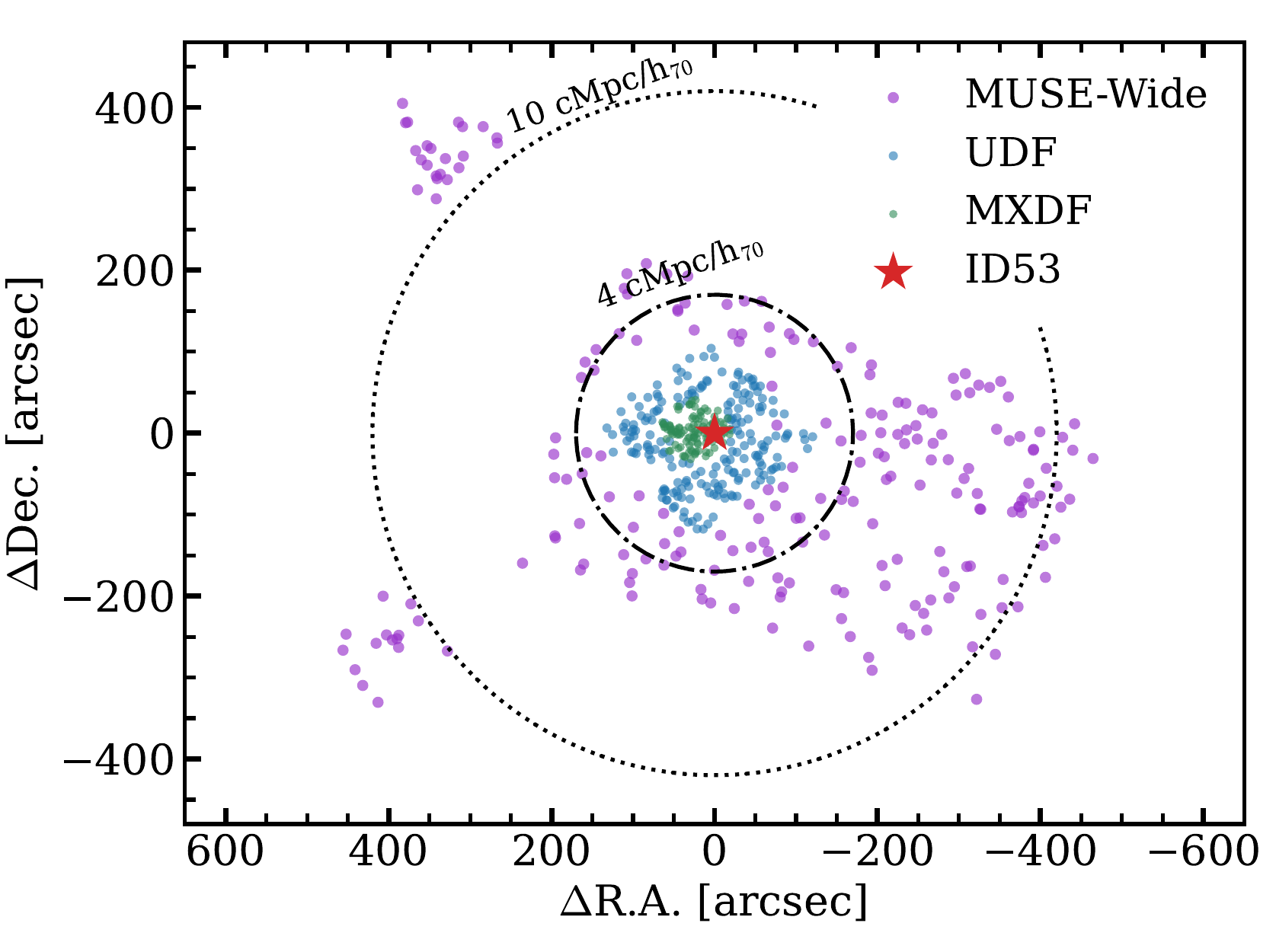}\vspace{-0.5cm}
    \caption{The spatial distribution of galaxies at $z=3.95-4.75$ in the foreground of ID53. Galaxies detected in the 1hr MUSE Wide footprint are shown in purple, those in the 10-31hr UDF region are shown blue and the galaxies in the deepest 140 hr MXDF region are shown in green. The coordinates are with respect to the background galaxy (R.A. = 03:32:37.95, Dec. = -27:47:10.94 in the J2000 reference frame) which is approximately in the center of the $3’\times3’$ UDF region. The two dashed-dotted and dotted circles illustrate impact parameters of 4 and 10 cMpc/h$_{70}$, respectively.}
    \label{fig:RADEC}
\end{figure}

The background galaxy `ID53' (ID number in the MUSE UDF catalog; \citealt{Bacon17}) is the continuum-brightest galaxy at $z>3$ in the deepest 140 hr MXDF coverage of the MUSE surveys in the Extended Chandra Deep Field South (see Fig. $\ref{fig:RADEC}$). It has an L$^{\star}$ UV luminosity ($m=25.2$) and lies at a redshift of $z=4.7745$ (measured through fine-structure ISM lines and consistent with the redshift from fitting high resolution stellar population models; \citealt{Matthee22}). As detailed in \cite{Matthee22}, the 1D spectrum is extracted using an optimal weighting based on its slightly extended continuum spatial profile at the MUSE resolution of $\approx0.5$''. The typical signal-to-noise ratio (SNR) of the flux density in the wavelength region of interest for measuring the Ly$\alpha$ forest ($\lambda\approx600-700$ nm) is 5 per resolution element ($\approx2.5$ {\AA}, see \citealt{Bacon22} for details), up to 20 for the regions with highest transmission. The S/N is higher in the wavelength regions redder than the Lyman break that were used to fit the spectral energy distribution (SED) of the galaxy.

In \cite{Matthee22}, we showed that the detailed rest-frame UV ($\lambda_0=1220-1600$ {\AA}) spectrum and its rest-frame optical photometry measured with {\it HST}/WFC3 and {\it Spitzer}/IRAC can be well described by a combination of star-light described by BPASS stellar population models \citep{BPASS} with a \cite{Chabrier03} initial mass function and a flexible star formation history. The star formation history was varied between a single burst and a continuous age distribution and the stellar metallicity was varied as well. The stellar SED models were attenuated by a uniform dust screen (with a \citealt{Reddy16a} attenuation law. The Ly$\alpha$ forest region was not included in the fit due to the unknown and stochastic impact of the IGM. The SED models well-match the general shape of the spectrum and the strength of metal sensitive wind features such as the NV P Cygni feature with young ages (log$_{10}$(age/yr)=6.5-7.6) and low metallicities ([Z/H]=-2.15 to -1.15). The low age and metallicity imply that the spectrum is relatively free of (strong) stellar absorption lines (see also \citealt{Cullen20}). This minimizes uncertainties in the identification or strength of Ly$\alpha$ forest features due to contamination or overlap with photospheric absorption. In our estimate of the Ly$\alpha$ transmission, we propagate the uncertainty in the intrinsic spectrum using all models explored in \cite{Matthee22} that have a $\Delta \chi^2_{\rm reduced}<1$ from the best-fit model of the rest-frame UV spectrum. These model uncertainties therefore also include further variations in the star formation histories, initial mass function and the inclusion of binary stars or not, in addition to statistical uncertainties. We notice that the uncertainty in the SED models has a few local maxima around the position of metal absorption lines such as CIII$_{\lambda 1175}$. We mask these regions in our analysis.

\begin{figure*}
    \centering
    \includegraphics[width=17.9cm]{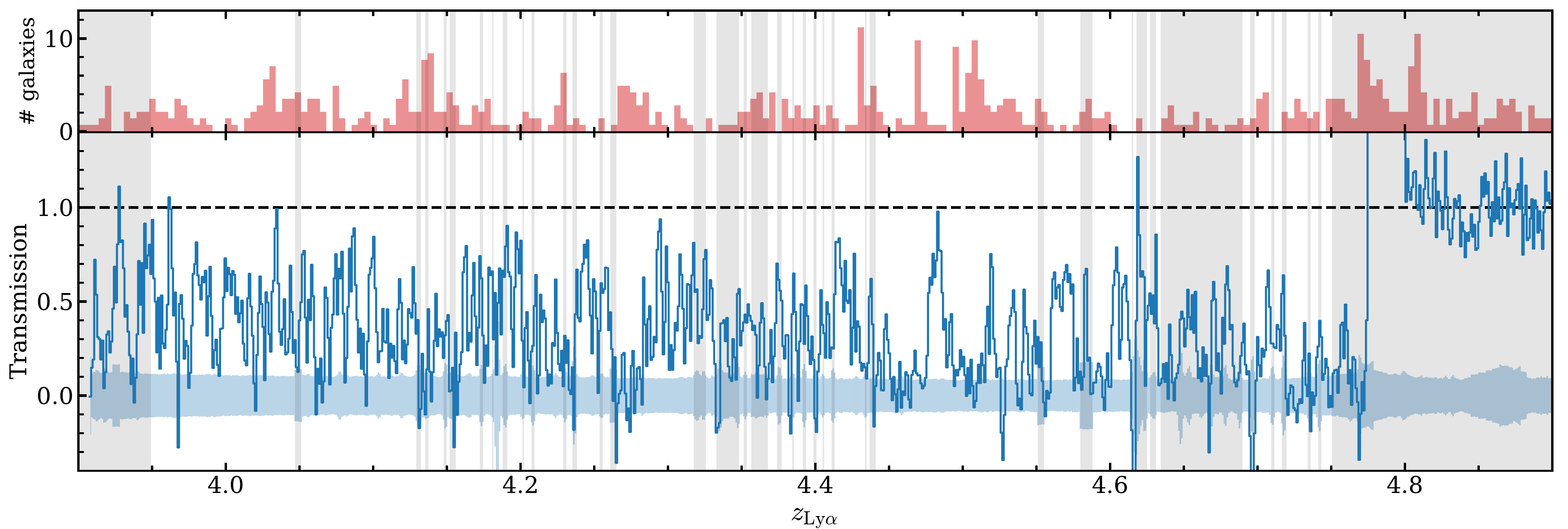}\vspace{-0.5cm}
    \caption{The HI Ly$\alpha$ transmission in the spectrum of background galaxy ID53 at $z=4.774$ (blue, bottom) compared to the foreground galaxy redshift distribution within the full field of view of MUSE surveys (red, top; the maximum impact parameter is 14 cMpc/h$_{70}$). Grey stripes highlight the regions masked due to skylines, relatively large SED model uncertainties (which are larger at metallicity-sensitive features), the locations of possible interstellar absorption lines and vicinity to ID53. Masked data were not included in measurements of the transmission or the cross-correlation between transmission and galaxy properties. The blue shaded region shows the uncertainty on the transmission which propagates both uncertainties on the intrinsic SED of ID53 and the measurement errors of the spectrum.} 
    \label{fig:spectrum}
\end{figure*}

The HI Ly$\alpha$ transmission, $T$, measured in the spectrum of our background galaxy is defined as 
\begin{equation}
    T = F_{\rm obs}/F_{\rm model},
\end{equation}
where $F_{\rm obs}$ is the observed flux and $F_{\rm model}$ the modeled galaxy spectrum before IGM absorption as described above. The bottom panel of Fig. $\ref{fig:spectrum}$ shows the HI Ly$\alpha$ transmission in our sight-line as a function of redshift. The uncertainties in the transmission account for both measurement uncertainties and the uncertainties in the SED fit. The lower redshift limit for our Ly$\alpha$ transmission measurements is $z=3.95$, driven by the observed wavelength range that is blocked because of the laser used for the ground-layer adaptive optics corrections and. The upper redshift limit is $z=4.72$ in order to avoid any line of sight effects associated to the presence of the background galaxy itself. This line of sight velocity distance corresponds to $>2800$ km s$^{-1}$, or $\gtrsim22$ cMpc/h$_{70}$. We mask redshifts regions that are impacted by atmospheric skylines, possible interstellar (FeII, SiII, SiIII) and stellar absorption lines from the background galaxy, but we note that including these data would have little impact on our results.


\subsection{Foreground galaxy sample} \label{sec:foreground}
The foreground galaxies that we investigate are spectroscopically confirmed galaxies identified by MUSE surveys in the Extended Chandra Deep Field South. The redshifts of the galaxies are determined using the bright Ly$\alpha$ line (i.e. these are Ly$\alpha$ emitters; LAEs). The sample originates from a combination of two catalogues: the DR2 UDF catalog that combines data from MOSAIC (10 hour), UDF-10 (30 hour) and MXDF (140 hour; see \citealt{Bacon22}), and the MUSE Wide (MW, 1 hour) catalog \citep{Herenz17,Urrutia19}, which is shallower, but covers a larger area (see Fig. $\ref{fig:RADEC}$). The background galaxy is located roughly in the middle of the UDF. We note that a significant region of the UDF-10 is superseded by the deeper MXDF survey. In total there are 291 galaxies in the UDF field and 213 galaxies in the MW field between $z=3.95-4.72$.

\begin{figure}
    \centering
    \includegraphics[width=8.5cm]{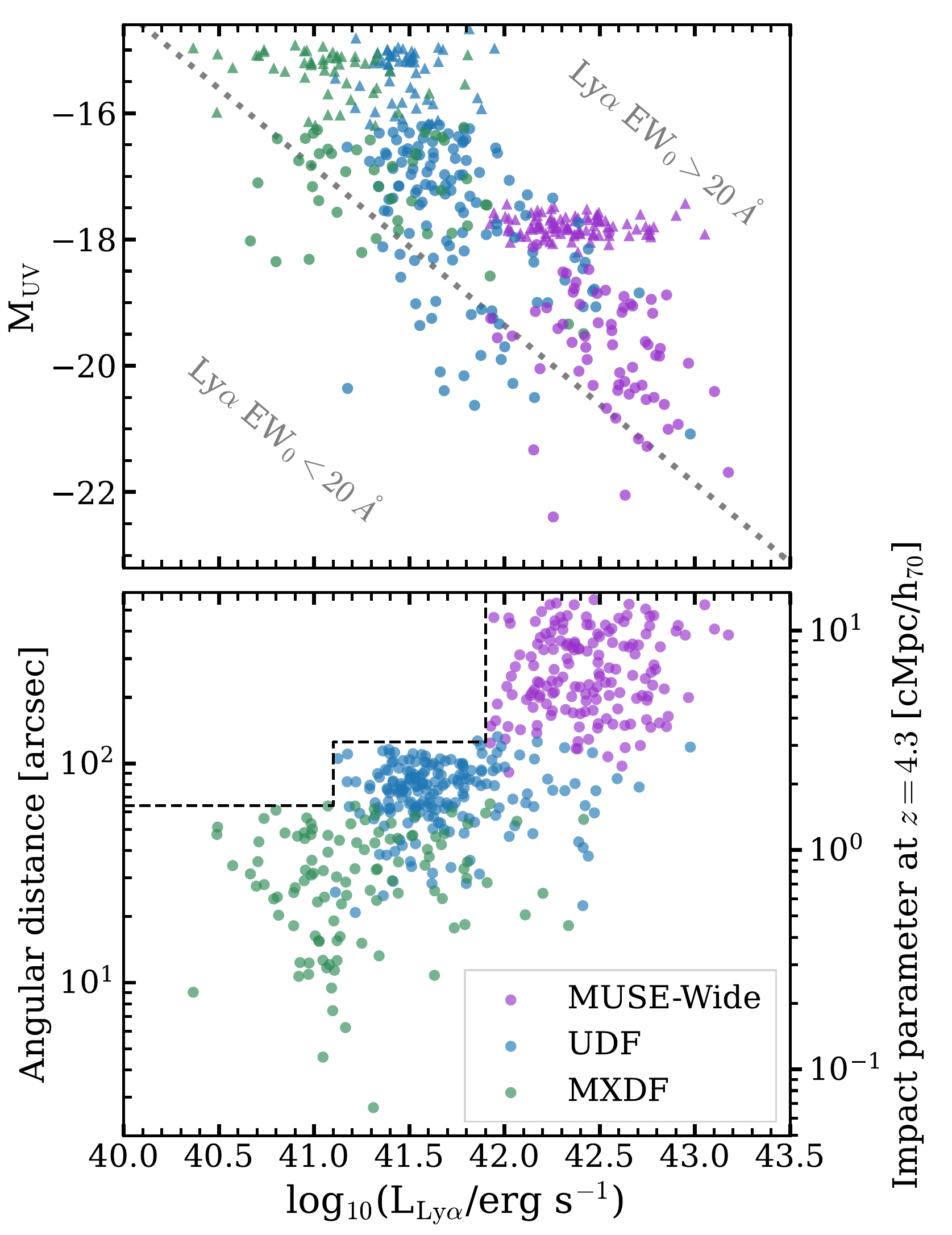}\vspace{-0.5cm}
    \caption{The Ly$\alpha$ and UV luminosities (top) and impact parameters (bottom) of the foreground galaxy sample. The limiting sensitivity of the {\it HST} data corresponds to M$_{\rm UV} \approx -15.5$ at $z\approx4.3$. Non-detections in the {\it HST} data are shown with upward pointing triangles. In the top panel, we illustrate the demarcation of galaxies with a rest-frame Ly$\alpha$ EW above and below 20 \AA assuming a UV slope of $\beta=-2$. In the bottom panel, the dashed lines illustrate the limiting sensitivity at $z=4.3$ and field of view of the MXDF, UDF and MW surveys, respectively. } 
    \label{fig:stats}
\end{figure}

Fig. $\ref{fig:stats}$ shows the Ly$\alpha$ and UV luminosities (top panel) and impact parameters (bottom panel) of the sample.  The Ly$\alpha$ luminosities are measured from the MUSE data and UV luminosities are determined using tabulated {\it HST}/ACS F850LP magnitudes \citep{Guo13}, where the limiting magnitude corresponds to M$_{\rm UV}\approx-15$. A significant fraction of the Ly$\alpha$-selected galaxies are undetected in the {\it HST} data, even down to a magnitude $\approx30$ (see \citealt{Maseda18}), explaining the pile-up in the top-left corner in Fig. $\ref{fig:stats}$. Our sample spans a large dynamic range of 0.004-2.5 L$^{\star}_{\rm UV}$ and a factor $\approx1000$ in Ly$\alpha$ luminosity, with a median M$_{\rm UV}=-17.9$, L$_{\rm Ly\alpha} =6.5\times10^{41}$ erg s$^{-1}$ and rest-frame Ly$\alpha$ EW $\approx50$ {\AA}. Most galaxies in our sample (92 \%) are Ly$\alpha$ emitters with a rest-frame Ly$\alpha$ equivalent width (EW) above 20 {\AA}. The median impact parameter is 93 arcsec which corresponds to 620 pkpc or 2.3 cMpc/h$_{70}$ at $z=4.3$. This is roughly 50 times the virial radius of halos with mass $\approx10^{10-11}$ M$_{\odot}$ which these galaxies are expected to reside in \citep{HerreroAlonso2022} at $z\approx4$. This means that most of the gas we observe in absorption is not physically related to the detected galaxies at an individual level, but rather traces the large-scale distribution where the galaxies reside in. The distribution of impact parameters depends on Ly$\alpha$ luminosity due to the tiered survey design. In Section $\ref{sec:CC}$, we adopt the limiting luminosities of our samples to control for any impact this design has on the cross-correlation .

Due to resonant scattering in the interstellar medium, the emerging Ly$\alpha$ line profile from galaxies is typically asymmetric \citep[e.g.][]{Wehrse79,Verhamme06,Gronke17,Dijkstra19} with a dominant peak that is redshifted with respect to the systemic redshift \citep[e.g.][]{Pettini98,McLinden11,Rakic11,Erb14,Trainor15,Cassata2020,Muzahid20,Matthee21}. The velocity offset is typically measured to be $\approx+200$ km s$^{-1}$ for Ly$\alpha$-selected galaxies and it depends on Ly$\alpha$ EW \citep{Adelberger2003,Nakajima18} and the Ly$\alpha$ line width \citep{Verhamme18}. As described in \cite{Bacon22}, the UDF-DR2 catalog provides an estimate of the systemic redshift of the galaxies based on the empirical relation between Ly$\alpha$ line width and the velocity offset from \cite{Verhamme18}, in case the systemic redshift could not be measured directly (which is the case for almost all galaxies in our sample). The intrinsic uncertainty on this method is $\approx90$ km s$^{-1}$ \citep{Verhamme18}, which corresponds to about 1 cMpc/h$_{70}$ at $z=4.3$ under the Hubble flow. The MW catalog only reports Ly$\alpha$ redshifts for the galaxies in our sample \citep{Herenz17}. In our cross-correlation analysis, we use refined estimates of the systemic redshifts based on the stacked HI absorption data around galaxies \citep[e.g.][]{Rakic11,Muzahid20} as identified and discussed in detail in \S $\ref{sec:CC}$.

\subsection{3D galaxy density field} \label{sec:densmethod}
The MUSE UDF region is the field with the highest sky-density of spectroscopic redshifts \citep{Inami17}. We use this high density of redshifts to construct a three dimensional map of the galaxy density around the sight-line to the galaxy ID53, enabling us to directly compare density and HI transmission. 

The density field is constructed using a kernel density estimator as follows. We create a cube with a grid with cells of size 0.2 cMpc/h$_{70}$ (ignoring any impact of motions along the line of sight on galaxy redshifts and using self-refined estimates of systemic redshift detailed in \S $\ref{sec:CC}$) spanning $\pm5$ cMpc in the transverse distance from ID53 and $z=3.8-4.9$ in the line of sight direction. We limit ourselves to investigating the density field in the UDF region in order not to be impacted too much by varying survey depth. The grid extends beyond the UDF region such that boundary effects do not impact the smoothing (see below). In order to overcome biases due to the variations in the depth within the central UDF region (see Fig. $\ref{fig:stats}$) we also limit the galaxy sample to those with a Ly$\alpha$ luminosity above $10^{41.1}$ erg s$^{-1}$ which is roughly the detection limit of the UDF region \citep{Drake17b}, see Fig. $\ref{fig:stats}$, and for which their luminosity function can be used. We count the number of galaxies in each cell, ignore cells that correspond to masked volumes (illustrated in Fig. $\ref{fig:spectrum}$) and we derive the excess density:
\begin{equation}
\delta_{\rm gal} = N_{\rm gal, observed}/N_{\rm gal, expected} -1,
\end{equation}
where $\delta_{\rm gal}$ is the excess density, $ N_{\rm gal, observed}$ the observed number of galaxies per cell and $N_{\rm gal, expected}$ the expected number of galaxies per cell based on the Ly$\alpha$ luminosity function at $z=4-5$ from \cite{Drake17b}, which agrees with the mean density that we measure in this line of sight. We assume that there is no redshift evolution in the luminosity function over $z=3.8-4.9$ \citep[e.g.][]{Sobral18,Herenz19}. Finally, we smooth the density cube with a spherical kernel of 2 cMpc/h$_{70}$ (see e.g. \citealt{Darvish17}). We have varied the smoothing kernel by a factor of two and find little impact on our results.

\section{THE EVOLVING HI OPACITY IN THE MXDF} \label{sec:tauevo}
We use our transmission data to measure the evolution of the HI opacity in terms of an effective optical depth, $\tau_{\rm eff}(z) \equiv -\ln(T)$, where $T$ is defined as in Eq. $\ref{eq:T}$. Following typical measurements in quasar sight-lines \citep[e.g.][]{Fan06,Becker15}, we bin our data in subsets of 50 cMpc/h$_{70}$ along the line of sight. The specific positions of the bin edges are chosen to maximise the number of bins in our data and we mask various data-points around skylines and more uncertain estimates of the intrinsinc emission as detailed in \S $\ref{sec:methods}$. The highest redshift bin ($z=4.71\pm0.06$) may be affected by the presence of the bright background galaxy at $z=4.7745$ that lies in an over-density and should therefore be interpreted with caution. Table $\ref{table:opacity}$ lists our measurements. The 1$\sigma$ uncertainties on these measurements have been derived by propagating the errors on the transmission data.

\begin{figure}
    \centering
    \includegraphics[width=8.5cm]{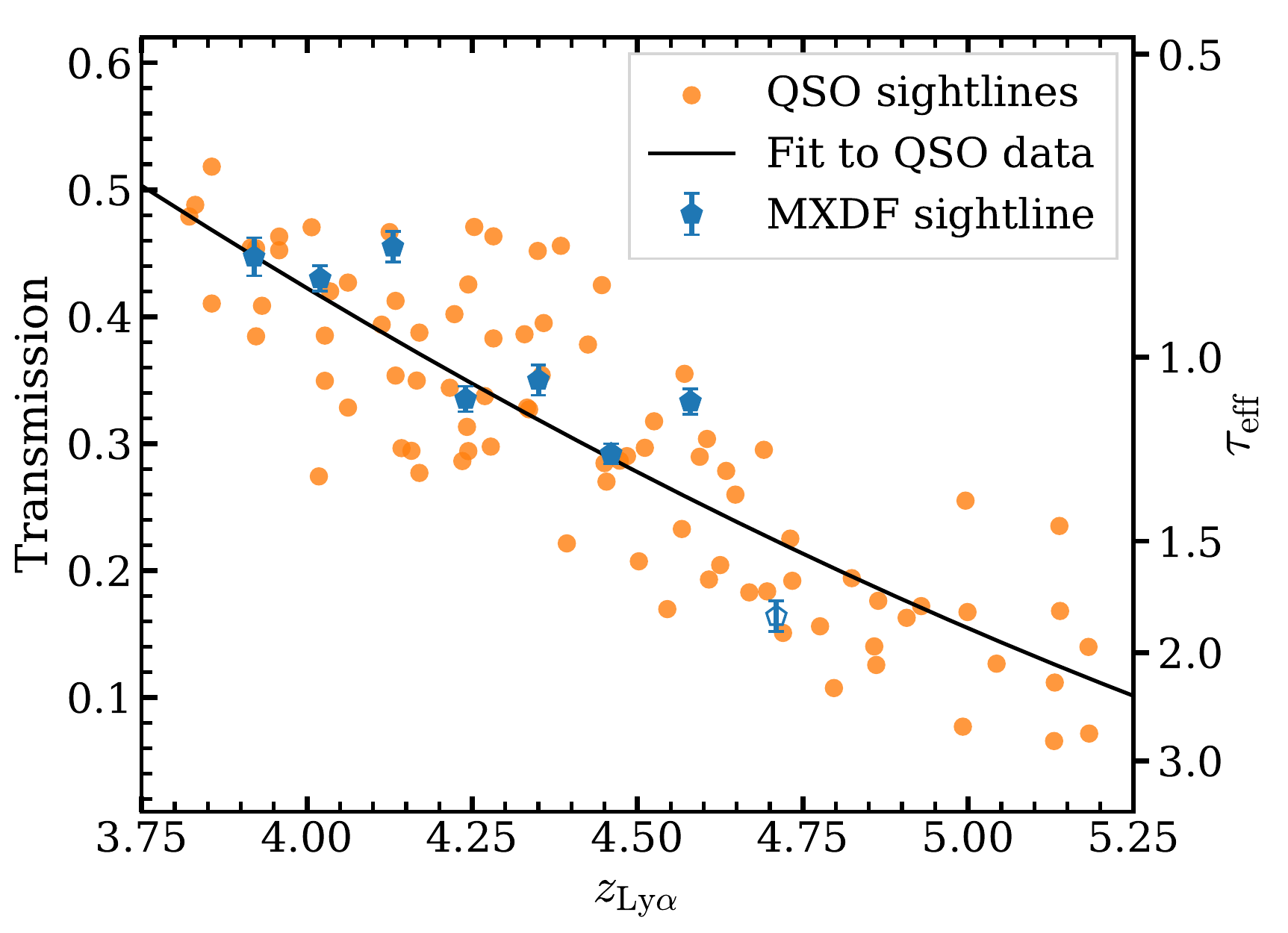}\vspace{-0.5cm}
    \caption{HI Ly$\alpha$ transmission (measured in bins of 50 cMpc/h$_{70}$; corresponding $\tau_{\rm eff}$ on the right) versus redshift. Orange data-points are measured in quasar sight-lines \citep{Becker15}. The blue hexagons show our measurements in the sight-line to the galaxy ID53 in the MXDF. Our highest-redshift data-point (open symbol) may be impacted by the proximity to the background galaxy (and its associated over-density, see Fig. $\ref{fig:spectrum}$). The polynomial fit to the data in quasar sight-lines is shown as a black line (Equation $\ref{eq:T}$).} 
    \label{fig:transmission}
\end{figure}

As shown in Fig. $\ref{fig:transmission}$, the normalisation and redshift evolution of the HI opacity in the MXDF averaged over these 50 cMpc/h$_{70}$ scales is in good agreement with the opacity measured in quasar sight-lines (from \citealt{Becker15}). This consistency yields a first and useful verification of completely independent methods for estimating the HI transmission at high redshift that are subject to different systematics. While our method relies on dust attenuation curves \citep{Reddy16a} and (theoretical) stellar population models that have been shown to provide good fits to continuum spectra of high-redshift galaxies \citep[e.g.][]{Steidel16,Cullen19,Matthee22}, quasar transmission measurements at these redshifts mostly rely on spline fits of the continuum that are subject to renormalisation corrections (as even voids have a relatively high opacity and the precise continuum level is therefore challenging to determine; \citealt{Becker11}). We note that the MXDF measurements scatter lies well within the range sampled by quasar sight-lines. Our measurement errors are too small to explain the scatter within our own data, suggesting a physical origin.

\begin{table}
\centering
\caption{The HI Ly$\alpha$ transmission $T$ and optical depth $\tau_{\rm eff}$ measured as a function of redshift, in bins of 50 cMpc/h$_{70}$ in the sight-line to galaxy ID53 in the MXDF.}
\label{table:opacity}
\begin{tabular}{lcc}
\hline
\multicolumn{1}{l}{$z$} &
\multicolumn{1}{c}{$T$} &
\multicolumn{1}{c}{$\tau_{\rm eff}$}\\
\hline
\noalign{\smallskip}
3.92& $0.447\pm0.015$ & $0.806\pm0.034$ \\
4.02 & $0.430\pm0.010$ & $0.843\pm0.023$ \\
4.13 & $0.455\pm0.012$ & $0.787\pm0.026$ \\
4.24 & $0.335\pm0.010$ & $1.093\pm0.031$ \\
4.35 & $0.350\pm0.012$ & $1.051\pm0.034$ \\
4.46 & $0.292\pm0.008$ & $1.230\pm0.026$ \\
4.58 & $0.333\pm0.010$ & $1.099\pm0.029$\\
4.71 & $0.164\pm0.012$ & $1.810\pm0.071$ \\ \hline
\noalign{\smallskip} 
\end{tabular}
\end{table}

In the remainder of the paper, we investigate {\it excess transmission} with respect to the average transmission at a certain redshift (i.e. the intergalactic mean). As the typical transmission evolves significantly with redshift at $z>4$,  we fit the simple polynomial shown in Fig. $\ref{fig:transmission}$ as a baseline. In order not to bias our results, we only use the quasar data to derive the average transmission. While we only study the $z=3.95-4.72$ redshift interval, we fit over the longer $z=3.8-5.2$ baseline. The best-fit polynomial is:

\begin{equation} \label{eq:T}
    \langle T(z)\rangle=2.372-0.663 z +0.0439z^2,
\end{equation}
where we note that the average transmission in the redshift range of interest is uncertain by $\approx0.03$ (i.e. about 10 \%). We measure an average transmission of 0.348 in the MXDF sight-line, which is 2 \% higher than the transmission expected based on this Equation (0.340) and thus well within the uncertainties.

\section{THE DENSITY - HI TRANSMISSION RELATION AT $z\approx4$} \label{sec:denstrans}
We now combine our transmission measurements with the estimated galaxy density (\S $\ref{sec:densmethod}$) along the sight-line and investigate whether and how the two are correlated. This approach allows us to probe the full dynamic range of over and under-dense regions, as opposed to the galaxy centric approach that we undertake in \S $\ref{sec:CC}$. In order to control for the strong redshift evolution of the transmission, we focus on the excess transmission $T/\langle T(z)\rangle -1$, where $\langle T(z)\rangle$ is described by Equation $\ref{eq:T}$. We do not include transmission datapoints that are more uncertain as discussed in \S $\ref{sec:backgroundgal}$ and illustrated in Fig. $\ref{fig:spectrum}$.

\begin{figure}
    \centering
    \includegraphics[width=8.5cm]{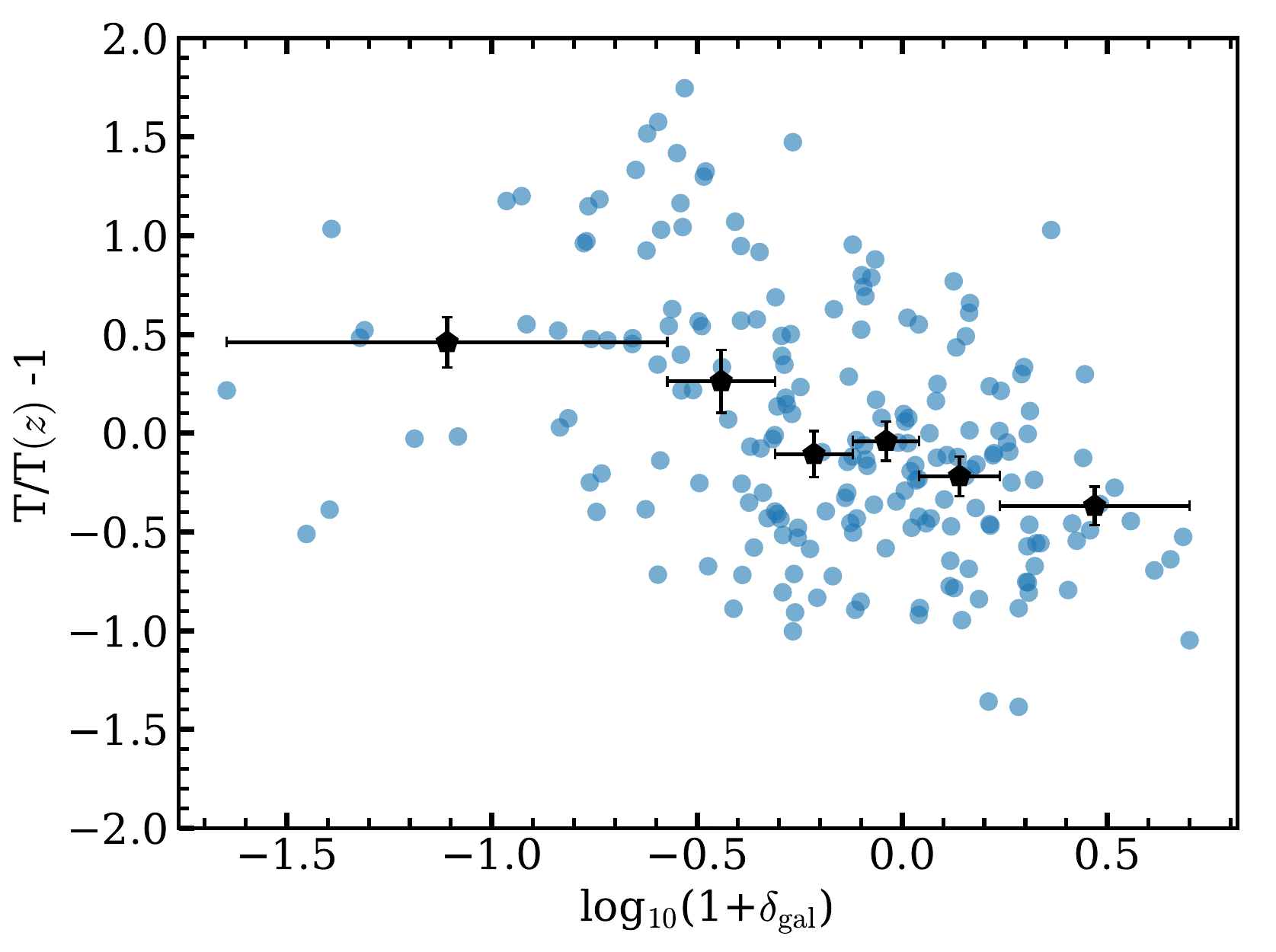} \vspace{-0.5cm}
    \caption{The relation between the galaxy density (smoothed over 2 cMpc/h$_{70}$ scales, see \S $\ref{sec:densmethod}$) and the HI Ly$\alpha$ transmission measured along the MXDF sight-line over $z=3.95-4.72$. Black data-points show binned averages, each averaging over the same number of data-points. We find an anti-correlation between galaxy density and transmission at 6.2$\sigma$ confidence. Over-dense regions are associated to regions with a low transmission, and vice versa. We note that individual data-points are relatively uncertain, explaining some values below -1.} 
    \label{fig:transdensity}
\end{figure}

Fig. $\ref{fig:transdensity}$ shows that we detect an anti-correlation between the excess transmission and galaxy over-density. Over-dense regions are associated with regions with a low transmission such that a region that is over-dense by a factor $\approx2$ on 2 cMpc/h$_{70}$ scales has a two times lower transmission than average. Under-dense regions likewise have a relatively high transmission. This is qualitatively consistent with results from \cite{Bielby20} in a shallower $z\approx4-5$ MUSE survey in a quasar field. In our data, the relation has a Spearman rank correlation coefficient of $r_s=-0.42$, which implies a $6.2\sigma$ significance considering the number of data-points. As a check, we have split our data into two redshift samples above and below $z=4.35$ and we find a similar trend in both sub-sets suggesting little redshift evolution within this range. The strength and significance of the anti-correlation depends on the kernel size over which the galaxy density is smoothed, ranging from $r_s=-0.34,-0.37,-0.42,-0.27,-0.22$ for kernel sizes of 0.5, 1, 2, 3 and 4 cMpc/h$_{70}$, respectively. The weakest anti-correlation (for 4 cMpc/h$_{70}$ kernels) is still detected at 3$\sigma$ significance. We note that redshift uncertainties of 90 km s$^{-1}$ (see \S $\ref{sec:foreground}$) correspond to uncertainties of $\approx 1$cMpc/h$_{70}$ in the line of sight direction assuming that the Hubble expansion dominates the apparent velocities, which likely yield a resolution floor of $\approx1$ cMpc/h$_{70}$ for the galaxy density map. It is therefore possible that the real correlation between the transmission and density could be stronger on scales $<1$ cMpc/h$_{70}$, but more accurate redshifts (i.e. systemic redshifts) are required to test this.

\section{EXCESS HI ABSORPTION AROUND GALAXIES} \label{sec:CC}
We use a stacking analysis of the Ly$\alpha$ absorption to study the properties of HI gas as a function of distance from foreground galaxies (either in the line of sight or projected direction, or combined). Because the absorption is measured in a single background spectrum, each piece of information (i.e. Ly$\alpha$ absorption signal) enters the stack multiple times, but at different projected or velocity distances depending on the galaxy used as a reference frame. We use the stack to validate -- and where needed refine -- Ly$\alpha$ redshift corrections that are available in the literature.

\subsection{Stacking methodology}
Averaged 1D renormalised transmission spectra and their uncertainties are obtained as follows. We re-normalise the transmission spectrum by the average transmission at a redshift $z$ (T($z$), Eq. $\ref{eq:T}$). We then shift the transmission spectrum to the rest-frame of the foreground galaxies, sample the shifted spectrum on a 80 km s$^{-1}$ grid and compute the stacked mean transmission averaging over the galaxies in the sample. This grid is coarser than the native MUSE sampling ($\approx50$ km s$^{-1}$), but similar to the full width half maximum of the line-spread function at the typical wavelength of the absorption spectrum and the redshift errors.

Our set-up of having $\approx300$ foreground galaxies and only a single background spectrum is quite different from typical surveys that use multiple quasar sight-lines \citep[e.g.][]{Bielby11,Rakic12,Muzahid21} or (stacks of) multiple background galaxies \citep{Steidel10,Chen20}. A consequence is that individual transmission data-points are often repeatedly included in stacks (at different velocities). Further, the clustering of the Ly$\alpha$ forest and galaxy populations induces relatively strongly correlated signal in the 1D spectrum on scales significantly larger than the line spread function. It is therefore challenging to properly estimate uncertainties on our stacked spectra. While surveys with multiple background sources typically estimate errors using bootstrap samples of their galaxy samples, we choose to use 1000 block-bootstrapped samples of the transmission data (with replacement) to account for such correlated noise in the spectral direction \citep[e.g.][]{Schaye2003,Loh08}. Bootstrapping the data in blocks of adjacent data-points leads to a higher noise estimate. While convergence is poor (noise estimates continue to increase with increasing block lengths; e.g. \citealt{Rollinde13}), we choose a block-length of 7 transmission bins (combined 560 km s$^{-1}$, or $\approx5$ cMpc/h$_{70}$ at the average redshift of the sample) as we find that this is the block length where the second derivative of the relation between block length and the noise becomes negative. For a block-length of 7, the noise estimate is about 26 \% higher compared to a block-length of 1. A block-length of 14 would have yielded 30 \% higher noise compared to standard. Besides the noise, we also use the bootstrap samples to assess biases in our transmission measurement. We find that the average excess transmission (at random velocity offsets with respect to galaxies) is 0.03, which is due to the entire sight-line having slightly higher transmission than our best-fit to the transmission in quasar sight-lines (see \S $\ref{sec:tauevo}$). We correct for this bias in our equivalent width measurements.

\subsection{Refined Ly$\alpha$ emission-based redshifts} \label{sec:zlya} 
Cross-correlation studies between HI absorption and galaxies require accurate systemic redshifts of galaxies \citep[e.g.][]{Adelberger05,Steidel10,Bielby11}. As detailed in \S $\ref{sec:foreground}$, the redshifts of the galaxies are measured using the Ly$\alpha$ emission-line, which is known to be typically redshifted with respect to the systemic redshift. An estimate of the systemic redshift is available for the majority of our galaxy sample (i.e. the 291 objects in the UDF+MXDF coverage at impact parameters $\lesssim3$ cMpc/h$_{70}$), see \cite{Bacon22}. This estimate is based on the theoretically motivated observed correlation between the offset of the red Ly$\alpha$ line and the systemic redshift and the line-width of the red Ly$\alpha$ line \citep{Verhamme18}. The average correction that is applied to obtain the systemic redshift is $-240$ km s$^{-1}$. The estimated uncertainty of this correction method is 90 km s$^{-1}$ \citep{Verhamme18}. 

\begin{figure}
    \centering
    \includegraphics[width=8.5cm]{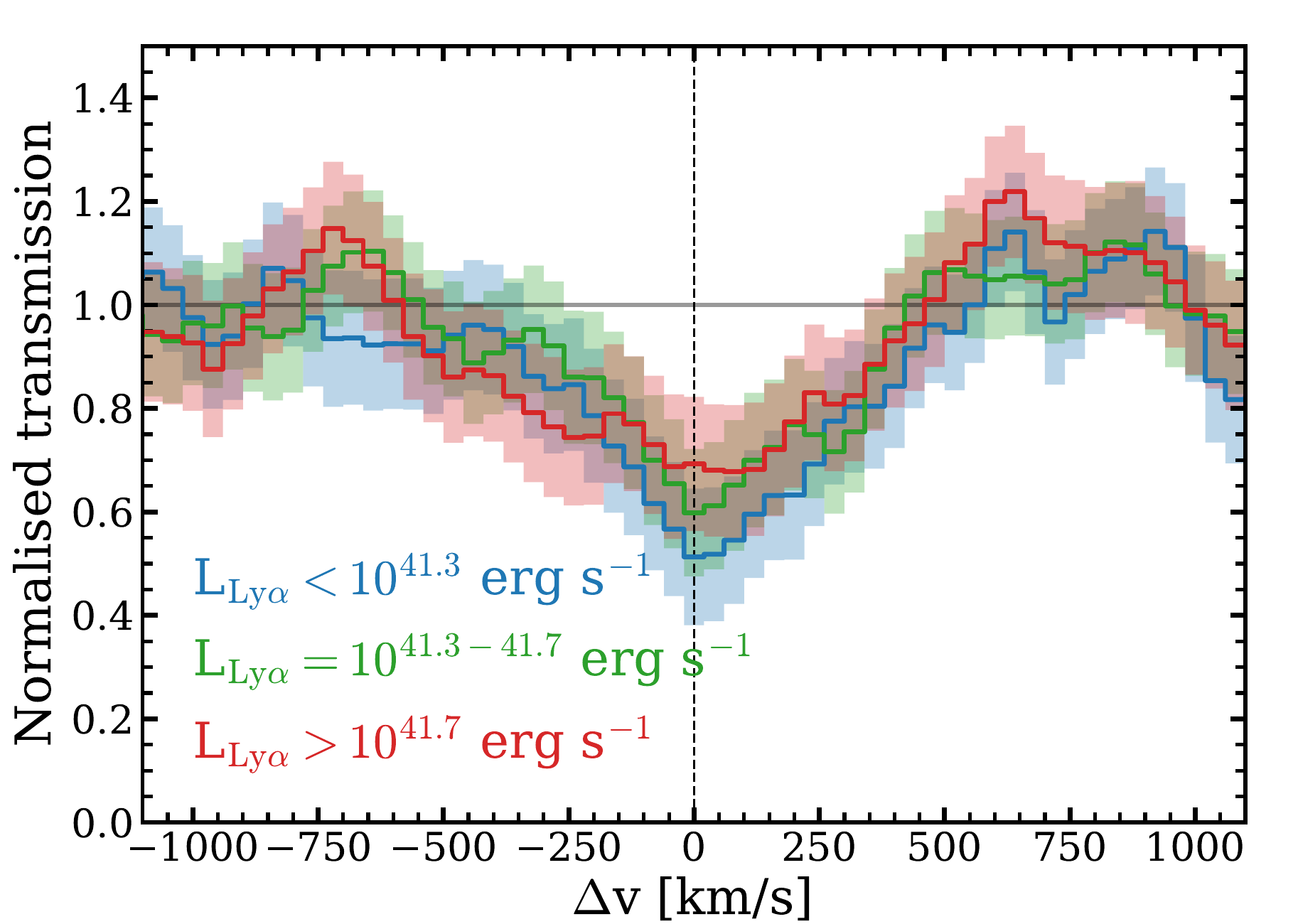} \vspace{-0.5cm}
    \caption{Stacked HI Ly$\alpha$ transmission spectrum for galaxies well within the UDF (impact parameter $\lesssim3$ cMpc/h$_{70}$) centered on the estimated systemic redshift based on Ly$\alpha$ line properties \citep{Bacon22}. The noise is indicated by the shading and estimated using block-bootstrapping. The galaxy sample is split by Ly$\alpha$ luminosity. } 
    \label{fig:Lyaoffset}
\end{figure}

Inspired by the approach from \cite{Rakic11} (see also \citealt{Muzahid20,Mukae20}), we can use our absorption line data to test the estimates of the systemic redshifts of our galaxy sample, and possibly refine them. The average HI absorption profile around our sample of LAEs should be symmetric if galaxy orientations are randomly distributed with respect to the background galaxy. Note that \cite{Momose21} argue that LAEs at $z\sim2$ may have anisotoropic distribution in the LOS direction relative to the HI gas. Such anisotropy may arise due to selection biases that impact the visibility of Ly$\alpha$ emission from galaxies, i.e. LAEs tend to locate in front of the HI gas relative to the observer, or have any specific preferred direction of peculiar motion. Such possible selection biases are ignored with our methodology. This method means that we shift the average systematic redshift until the absorption signal is symmetric around zero velocity. Compared to e.g. \cite{Rakic11}, this method is more challenging to perform with our single background spectrum due to the lower signal to noise and resolution compared to analyses using multiple quasar spectra, but as we show the method is still applicable. In Fig. $\ref{fig:Lyaoffset}$ we show the stacked HI absorption profile for galaxies within the UDF coverage, shifted to their estimated systemic redshift using the \cite{Verhamme18} method and split by Ly$\alpha$ luminosity, where each subset contains the same number of galaxies. We split by Ly$\alpha$ luminosity as this is the property that impacts the selection of the galaxies in our sample and that is most reliably and self-consistently measured (as opposed to any quantity related to the continuum luminosity). Errors and (small) renormalisations are estimated using block bootstrapping as described above. We quantify the asymmetry $a$ as $a=EW_{\rm blue} / EW_{\rm red}$, where the blue and red subscipts refer to the EWs integrated over $-500$ to 0 and 0 to +500 km s$^{-1}$ with respect to the estimated systemic redshift.

\begin{figure}
    \centering
    \includegraphics[width=8.5cm]{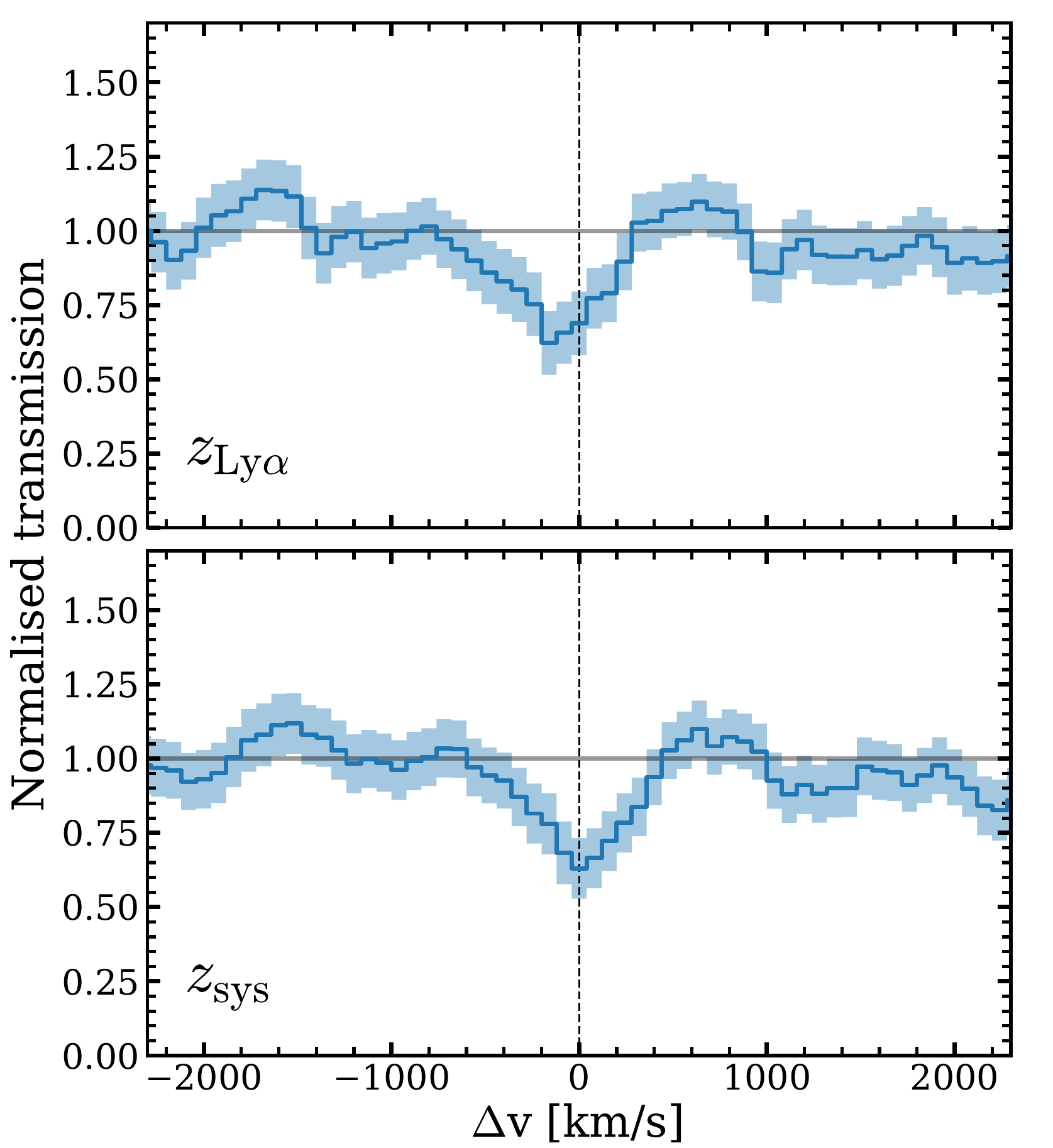} \vspace{-0.5cm}
    \caption{Stacked HI Ly$\alpha$ transmission spectrum for galaxies well within the UDF centered on the Ly$\alpha$ redshifts (top) and our refined estimates of the systemic reshifts (bottom). The noise is indicated by the shading and  estimated using block-bootstrapping.} 
    \label{fig:stacked_1D}
\end{figure}

As Fig. $\ref{fig:Lyaoffset}$ shows, we find that the estimated systemic redshifts are fairly accurate for the faintest LAEs in the sample (L$_{\rm Ly\alpha}<10^{41.3}$ erg s$^{-1}$) as the maximum absorption is detected at zero velocity and the absorption profile is fairly symmetric ($a=1.08\pm0.10$). For the two brighter subsets of LAEs, the estimated systemic redshifts are somewhat offset with respect to the velocity of maximum absorption (in particular the most luminous subset) and their profiles are also more asymmetric ($a=1.26\pm0.10$ and $a=1.12\pm0.10$, respectively). This suggests that the offsets between the Ly$\alpha$ line and the systemic redshifts are over-corrected by the \cite{Verhamme18} method for these brightest subsets. Indeed, we iteratively find that we can refine the systemic redshifts of these two subsets by applying the average of $+100$ and $+120$ km s$^{-1}$ corrections to the subsets with Ly$\alpha$ luminosities in the range $10^{41.3-41.7}$ erg s$^{-1}$ and above. The application of these offsets also leads to more symmetric stacked profiles for these subsets ($a=1.05\pm0.10$ and $a=0.97\pm0.10$, respectively). 

As a result, these refined velocity offsets for a sub-set of the sample imply that the average offset between the Ly$\alpha$ line and the systemic redshift is +170 km s$^{-1}$ for our sample under the assumption that LAE's $z_{\rm sys}$ is the same as the $z_{\rm sys}$ of the average absorbing HI gas in the associated large scale overdensity, with little dependence on luminosity. This average offset is in agreement with results at lower redshifts \citep[e.g.][]{Erb14,Matthee21}. Fig. $\ref{fig:stacked_1D}$ shows that the average absorption profile now centers on the systemic redshift and is symmetric ($a=1.01\pm0.05$), in particular compared to the average absorption profile when centering galaxies on their Ly$\alpha$ redshifts ($a=1.17\pm0.07$). The average rest-frame EW of the excess HI absorption (within $\pm500$ km s$^{-1}$ from the systemic reshift) is $0.74^{+0.39}_{-0.24}$ {\AA}. The excess absorption EW is fully consistent with the results at $z\approx3$ \citep{Chen20,Muzahid21} using a similar strategy, but with quasars as background sources. Based on these results, we also apply the average correction of 170 km s$^{-1}$ to the redshifts for the LAEs from the MUSE Wide catalog that we use at the largest impact parameters.

\subsection{How far out do we detect excess absorption?} \label{sec:LOStransverse}
\begin{figure}
    \centering
    \includegraphics[width=8.5cm]{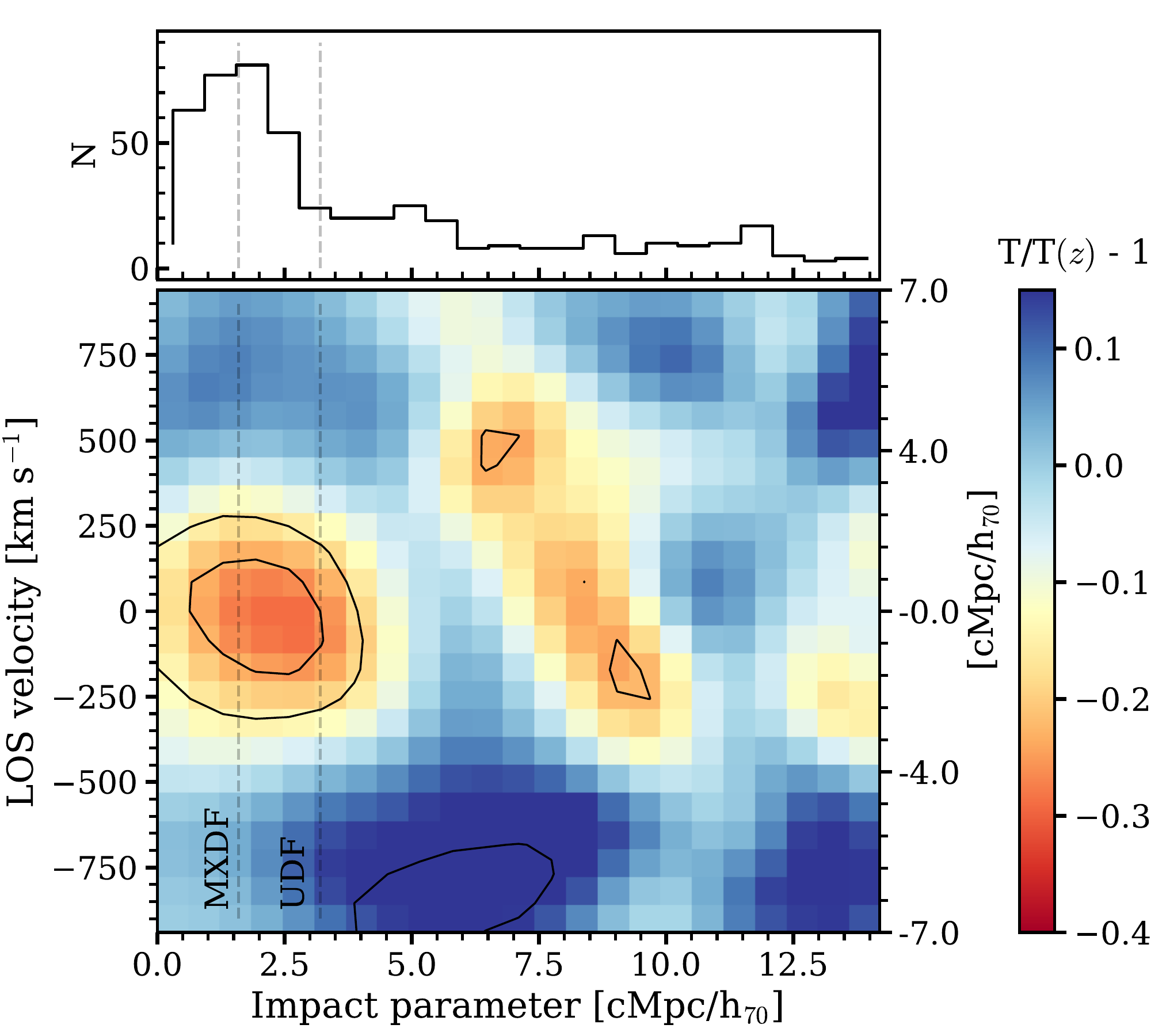} \vspace{-0.5cm}
    \caption{The excess transmission as a function of the impact parameter and the LOS velocity. Contours are at the 2,3 $\sigma$ level. In the top panel we show the histogram of impact parameters of the foreground galaxies. In the main panel, we highlight the maximum impact parameters corresponding to the deeper MXDF and UDF regions. The labels on the y-axis on the right translate the LOS velocity into a LOS distance for the average redshift of our sample ($z=4.3$) ignoring peculiar velocities.  }
    \label{fig:2Dplot}
\end{figure}

In order to address out to what scales excess absorption can be detected, we now create a mean stacked 2D map of the excess transmission as a function of impact parameter and line-of-sight (LOS) separation. We create a map with a grid-size of 80 km s$^{-1}$ in the LOS and 0.62 cMpc/h$_{70}$ (corresponding to $\approx80$ km s$^{-1}$ under the Hubble flow) in the transverse direction. Then, for each cell, we first select all galaxies in the corresponding range of impact parameters $b\pm\Delta b/2$ (where $\Delta b=0.62$ cMpc/h$_{70}$). Over the selected galaxies, we average the quantity $T(z+\Delta z)/\langle T(z)\rangle -1$, where $z$ is the systemic redshift of each galaxy based on Section $\ref{sec:zlya}$, $\langle T(z)\rangle$ the average transmission at redshift $z$ (Equation $\ref{eq:T}$) and $\Delta z$ corresponds to the grid binning of 80 km s$^{-1}$. We include all foreground galaxies in our sample that span impact parameters out to 14 cMpc/h$_{70}$. The uncertainty in each grid-cell is obtained through 1000 block-bootstrap samples in each impact parameter bin, similar to the 1D stacks described in \S $\ref{sec:zlya}$. The result is shown in Fig. $\ref{fig:2Dplot}$, which demonstrates that the excess absorption is significantly detected out to $\approx4$ cMpc/h$_{70}$ from galaxies. The excess absorption is similar in the transverse and LOS directions. 

Due to our survey design, the number of foreground galaxies is a strong function of impact parameter: more than 50 per impact parameter bin within impact parameters of $\approx4$ cMpc/h$_{70}$, but only $\approx10-20$ per bin at larger impact parameters. As a consequence, the uncertainty on the excess transmission is a strong function of impact parameter. Due to cosmological isotropy considerations, any real excess signal should be symmetric in the positive and negative LOS velocities. The indicative excess transmission/absorption in various regions at $\approx5-10$ cMpc/h$_{70}$ (Fig. $\ref{fig:2Dplot}$) is therefore implausible, and is likely due to the cosmic variance of a single sight-line. We note that we have verified that our results are not strongly impacted by the large variation in survey depth as a function of impact parameter due to the change from the UDF to the MUSE Wide region at $\sim4$ cMpc/h$_{70}$. If we limit our sample to galaxies with Ly$\alpha$ luminosities $>10^{42}$ erg s$^{-1}$, which are detectable across the entire field of view, we still find significant excess absorption at distances $\lesssim4$ cMpc/h$_{70}$ from galaxies. As illustrated in Fig. $\ref{fig:RADEC}$, scales of $\approx4$ cMpc/h$_{70}$ are already larger than the size of the UDF, demonstrating the value of the larger area of the MUSE Wide data.

Fig. $\ref{fig:2Dplot}$ also suggests that there is somewhat less excess absorption at the smallest impact parameters (corresponding to $\lesssim150-200$ pkpc). While the number of galaxies with these impact parameters is lower, leading to higher uncertainties, we speculate that the lowered excess absorption could partly be due the contribution from a proximity effect around galaxies \citep[e.g.][]{Kashino22}, emission-infilling from extended Ly$\alpha$ halos around galaxies \citep[e.g.][]{Wisotzki2018,Chen20}, or around their clustered satellites \citep{Kikuchihara22}.

\begin{figure}
    \centering
    \includegraphics[width=8.5cm]{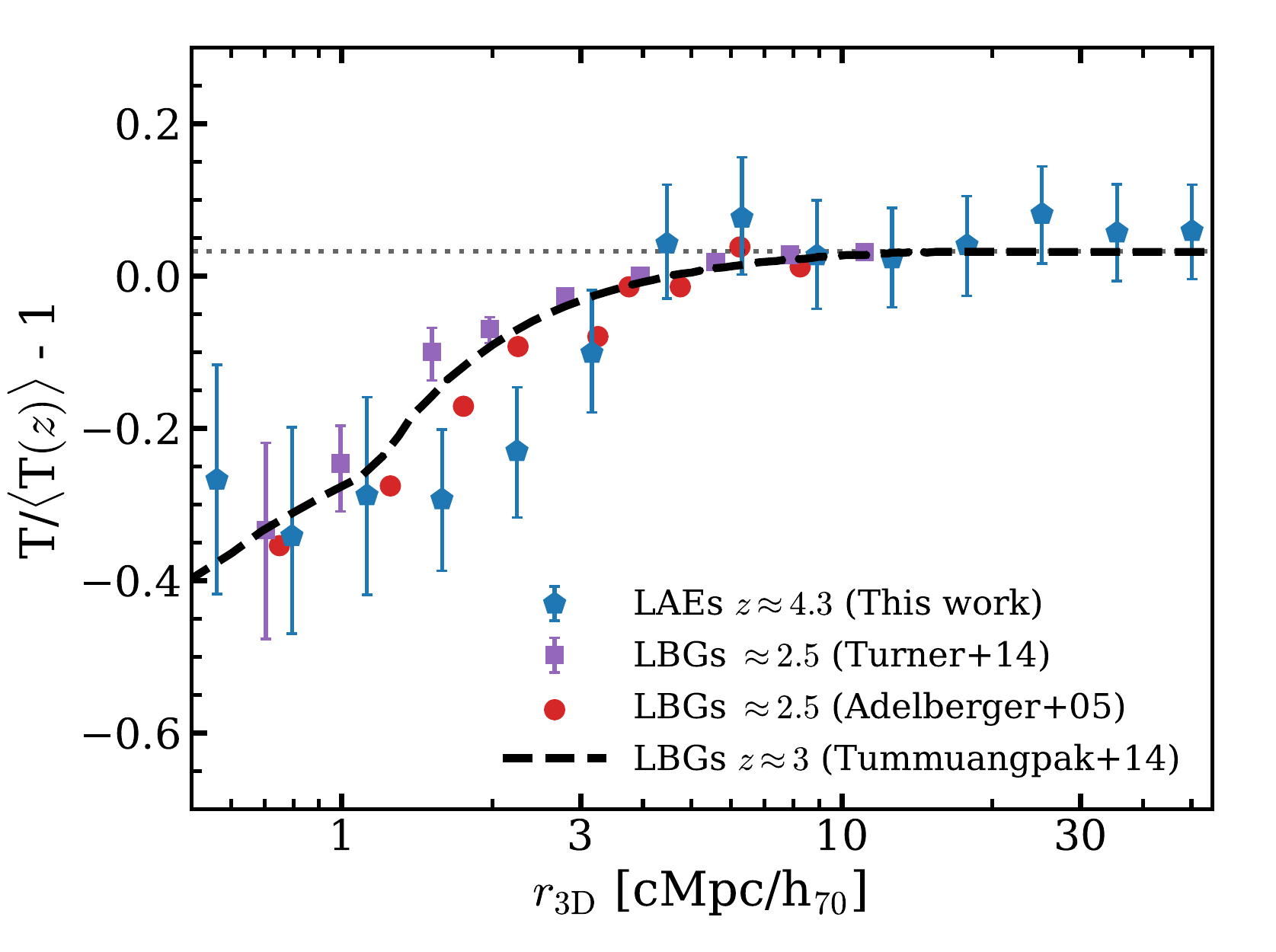} \vspace{-0.5cm}
    \caption{The excess HI transmission as a function of 3D distance for our full sample of foreground galaxies. Excess absorption is detected out to 4 cMpc/h$_{70}$ from galaxies. Uncertainties at small distances are large due to the limited number of sources and susceptibility to redshift errors. We do not subtract the slight bias in our excess transmission and illustrate it with a dotted line. The black dashed line (corrected for this bias) shows the typical excess transmission around LBGs at $z\approx3$ \citep{Tummuangpak14}, whereas purple squares and red dots show other measurements (from \citealt{Turner14} and \citealt{Adelberger05}, respectively) in LBGs at $z\approx2.5$ corroborating this curve.}
    \label{fig:3D}
\end{figure}

The similarity of the dependence of the excess absorption on LOS and transverse distance suggests that we can safely ignore peculiar velocities with the resolution of our data. We therefore derive the relation between the excess transmission and 3D distance and show this in Fig. $\ref{fig:3D}$, where we use logarithmically spaced bins and measure the errors with bootstrapping as described above. Fig. $\ref{fig:3D}$ shows excess absorption is detected out to 4 cMpc/h$_{70}$ from galaxies and further illustrates the slight bias towards an excess transmission of 0.03 compared to our fit to the transmission in quasar spectra at these redshifts. The absorption profile is compared to measurements around UV-bright galaxies at $z\approx2-3$ \citep{Adelberger05,Tummuangpak14,Turner14}, which indicates that the profile appears somewhat shallower at high-redshift, with particularly stronger absorption at distances of $\approx2-3$ cMpc/h$_{70}$. It is unclear whether this finding can be fully ascribed to a genuine redshift evolution, as the considered samples also have other differences: while our LAEs have typical UV luminosities M$_{\rm UV}\approx-18$, the typical galaxy in the LBG samples is $\approx$30 times much more luminous with M$_{\rm UV}\approx-21.5$ \citep{Adelberger2003,Bielby11}. Moreover, while our shot noise is relatively low due to the large number of foreground galaxies, cosmic variance may be substantial in our single sight-line \citep[see e.g. the discussion in][]{Garel2016}, and our comparatively large redshift uncertainties have the possible effect to flatten the real profile on scales $\lesssim2$ cMpc/h$_{70}$ \citep[e.g.][]{Adelberger2003}.

In Appendix $\ref{app:tests}$, we show how the inhomogeneous nature of the survey (in terms of sensitivity and coverage of the foreground sample) impacts the excess transmission measurements. As we show, the detection of excess absorption within $\sim3$ cMpc/h$_{70}$ is robust to the gaps in the survey design as well as the varying sensitivity. A jackknife error estimate shows that there is some uncertainty in the relative transmission at large impact parameters ($>5$ cMpc/h$_{70}$). However, we note that relatively large 3D distances (e.g. in Fig. $\ref{fig:3D}$) are not only probed by high impact parameters, but also by large velocity offsets at small impact parameters where we have much larger statistics.

\subsection{Wider applicability and Future developments} \label{sec:future}
The wedding-cake layered mosaic of exposure times in the MUSE Wide, UDF and MXDF yields a high number of galaxies at small impact parameters around the objects for which background spectra are measured, which is greatly beneficial for various cross-correlation measurements. For example, there are (17, 59, 205) foreground -- background galaxy pairs (based on 505 unique galaxies) at impact parameters within (100, 200, 500) pkpc, respectively, which is more than in the KBSS survey \citep{Chen20}, which has (10, 26, 90) foreground -- background pairs (based on $\approx3000$ galaxies) at these separations. However, the relatively large number of galaxies at small impact parameters cannot be fully exploited for measurements of the excess absorption within $\lesssim$ 1 cMpc/h$_{70}$ when the systematic redshifts uncertainties are $\sim100$ km s$^{-1}$ due to the scatter in Ly$\alpha$ velocity offset corrections \citep[e.g.][]{Verhamme18}. This could be addressed with a complete spectroscopic redshift survey of the foreground population at $z\approx4$ with rest-frame optical lines, e.g. using {\it JWST}'s NIRCam grism mode that can cover H$\alpha$ and H$\beta$+[OIII] at these redshifts \citep[e.g.][]{Matthee23}, which would also mitigate any possible concerns of the Ly$\alpha$ line selection missing the most massive (and dusty) galaxies. It will be particularly important to investigate how well the galaxy over-density probed by Ly$\alpha$-selected samples (which are known to avoid the most massive star forming galaxies due to their low Ly$\alpha$ escape fraction; e.g. \citealt{Matthee16}) traces the galaxy density from other probes.

The position of the MXDF was not chosen for this particular measurement. In fact, while ID53 is the brightest galaxy at $z>3$ in the deepest MUSE coverage (including all areas observed by more than 30 hours; \citealt{Bacon22}), the region with 10 hr MUSE coverage contains three more galaxies at $z=4.7-5.8$ that are equally bright or brighter and are young galaxies similarly to ID53. The larger MUSE Wide region contains seven high-redshift galaxies brighter than ID53, including one with magnitude 23.3 at $z=4.84$. While we discuss these galaxies in more detail in Appendix $\ref{app:depths}$, we note here that, in addition to being UV bright, the best galaxies for tomography studies are those with a young age and a low dust attenuation similar to ID53, as this type of galaxies have a relatively featureless (blue) continuum in the Lyman-$\alpha$ forest region. 

By experimenting with the spectrum of ID53, we find that a S/N decrease of a factor of $\sim1.7$ would still yield a comfortable detection of the presence of excess absorption, suggesting the analysis could be extended to somewhat fainter sources. A specifically designed  deep IFU pointing (or observations with a multi-object spectrograph) could therefore technically already obtain transmission measurements in about three background galaxies within a single MUSE pointing, yielding valuable information of the variation of the Ly$\alpha$ transmission at small transverse distances. 

In the future era of the unprecedented spectroscopic sensitivity that the Extremely Large Telescopes will bring in the $\lambda\sim0.5-2$ micron regime, galaxy -- HI transmission cross-correlations using multiple closely separated sight-lines could feasibly be extended to higher redshifts $z\approx6$, where the Ly$\alpha$ fluctuations are substantial at the end stages of cosmic reionization \citep[e.g.][]{Bosman2022}, and may be able to target the weaker metal absorption lines such as CIV, which is also sensitive to the shape of the ionizing spectrum.

\section{SUMMARY} \label{sec:summary}
In this paper we have used deep spectroscopic data from MUSE surveys in the Extended Chandra Deep Field South to measure the Ly$\alpha$ IGM transmission and perform cross-correlation with the spectroscopically identified foreground galaxy density at $z=3.95-4.72$ in the spectrum of an L$^{\star}$ background galaxy at $z=4.77$, for the first time using a galaxy spectrum at such high redshift. This serves as a proof of concept for future studies where the use of galaxies as background source will become common to perform IGM tomography out to the epoch of reionization with a high spatial sampling.
Our results can be summarised as follows:

\begin{itemize}
    \item We measure the evolution of the effective HI IGM opacity $\tau_{\rm eff} = 0.8 - 1.8$ in bins of 50 cMpc/h$_{70}$ over the redshift interval $z=3.9-4.7$ in the MXDF. We show that these measurements are consistent with independent measurements in quasar sight-lines that have been performed with a different technique, providing a useful cross-validation of state-of-the art modelling of both galaxy and quasar continua at this redshift. (Fig. $\ref{fig:transmission}$).
    
    \item Thanks to the high sky density of galaxies with known  spectroscopic redshifts between $z\approx4-5$, we construct the galaxy density map along the Ly$\alpha$ transmission sight-line and we show that the galaxy density in the MXDF (smoothed with a spherical 2 cMpc/h$_{70}$ kernel) anti-correlates (at the 6$\sigma$ confidence level) with the excess Ly$\alpha$ transmission compared to the average cosmic transmission $\langle T(z)\rangle$. This measurement (unlike galaxy-centric stacks) reveals that under-densities are associated to regions with a higher transmission compared to the average. Our results confirm that density effects dominate the (excess) transmission over ionization effects out to at least $z\approx4.5$. (Fig. $\ref{fig:transdensity}$). 

    \item By stacking the Ly$\alpha$ transmission spectra centered on the redshifts of foreground galaxies and assuming that the HI absorption profile should center symmetrically around galaxy systemic redshifts, we show that -- on average -- Ly$\alpha$ emission-line redshifts are redshifted by +170 km s$^{-1}$ with respect to the systemic redshift, in good agreement with other measurements at slightly lower redshifts.

    \item Within 3 cMpc/h$_{70}$ from galaxies, the excess HI absorption around faint M$_{\rm UV}\approx-18$ galaxies at $z\approx4$ has a comparable strength as measured around brighter galaxies at $z\approx3$ (Fig. $\ref{fig:3D}$). Excess HI absorption around galaxies is detected out to 4 cMpc/h$_{70}$, similarly in the line-of-sight and the transverse directions. There is an indication that the excess absorption at $z\approx4$ is stronger at distances $\approx$ 2-3 cMpc compared to measurements at $z\sim2-3$, but given the various differences between studies it is challenging to pinpoint the origin of this difference, which may also simply be due to cosmic variance.
\end{itemize}

Our results demonstrate the feasibility of using extremely deep spectroscopy to measure the Ly$\alpha$ transmission in the spectra of relatively typical background galaxies at high-redshift, extending the redshift range to $z\approx5$. In particular, galaxies that are characterised by young ages, low metallicities and low dust attenuation have a relatively flat and bright UV continuum in the $\lambda_0=912-1216$ {\AA} range \citep[e.g.][]{Cullen19,Matthee22}, making these particularly useful for such measurements.
Determining the redshift and physical scales at which there is the transition from density to ionization effects that determine the cross-correlation signal between Ly$\alpha$ transmission and galaxies may help constraining the properties and distribution of ionizing sources at or just after the epoch of reionization \citep[e.g.][]{Garaldi22}. In a future dedicated survey, for example with MOSAIC on the Extremely Large Telescope, one could perform such measurement on a region on the sky with multiple closely separated background sources at $z\sim6-7$, such as a high-redshift quasar surrounded by relatively bright, young galaxies, that will yield multiple closely separated sight-lines. Identifying such regions is a task for in the meantime.

\section*{Acknowledgements}
We thank the referee for constructive comments that helped improving the paper.
Based on observations collected at the European Southern Observatory under ESO programme 1101.A-0127. Funded by the European Union (ERC, AGENTS,  101076224). Views and opinions expressed are however those of the author(s) only and do not necessarily reflect those of the European Union or the European Research Council. Neither the European Union nor the granting authority can be held responsible for them.
GP acknowledges support from the Netherlands Research School for Astronomy (Nederlandse Onderzoekschool Voor Astronomie, NOVA). JB acknowledges financial support from the Funda\c{c}\~ao para a Ci\^encia e a Tecnologia (FCT) through national funds PTDC/FIS-AST/4862/2020, work contract 2020.03379.CEECIND, and research grants UIDB/04434/2020 and UIDP/04434/2020. TU and LW acknowledge funding by the European Research Council through ERC-AdG SPECMAP-CGM, GA 101020943. TG is supported by the ERC Starting grant 757258 'TRIPLE’.

\section*{Data availability}
The MUSE data in the UDF-MXDF region underlying this work is available through the website https://amused.univ-lyon1.fr/. The catalog of galaxies in the MUSE Wide region is available through  https://musewide.aip.de/.
The transmission spectrum measured in this paper will be available online through the publisher.

\bibliographystyle{mnras}
\bibliography{MasterBiblio.bib}

\appendix
\section{Impact of the inhomogeneous survey design} \label{app:tests}

\begin{figure}
    \centering
    \includegraphics[width=8cm]{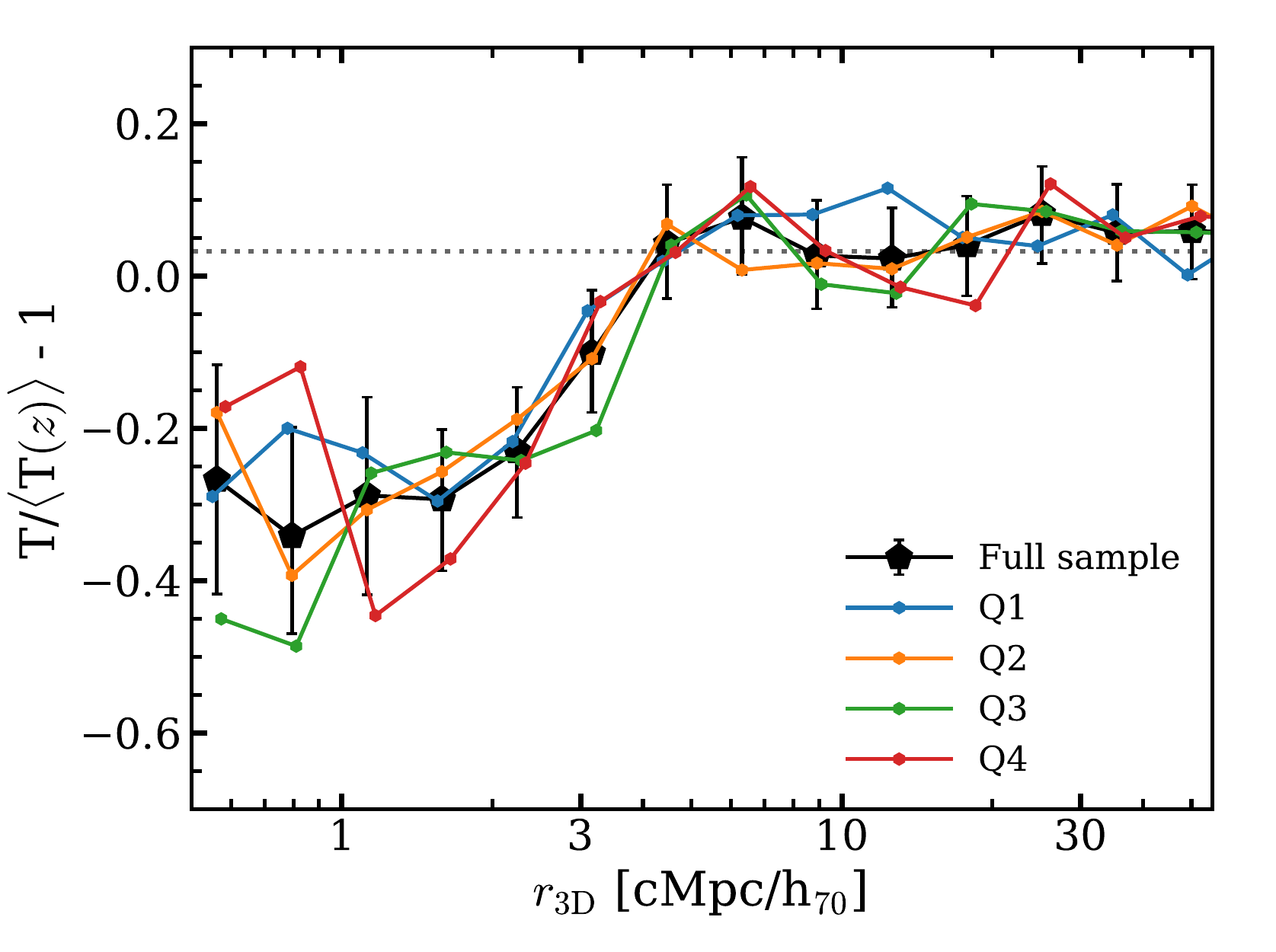}
    \caption{The excess HI transmission as a function of 3D distance to the foreground galaxies (as in Fig. $\ref{fig:3D}$) where the foreground sample is split in four quadrants (each with a different colour), compared to the full sample (black). }
    \label{fig:CC_jacks}
\end{figure}

\begin{figure*}
    \centering
    \begin{tabular}{cc}
        \includegraphics[width=7cm]{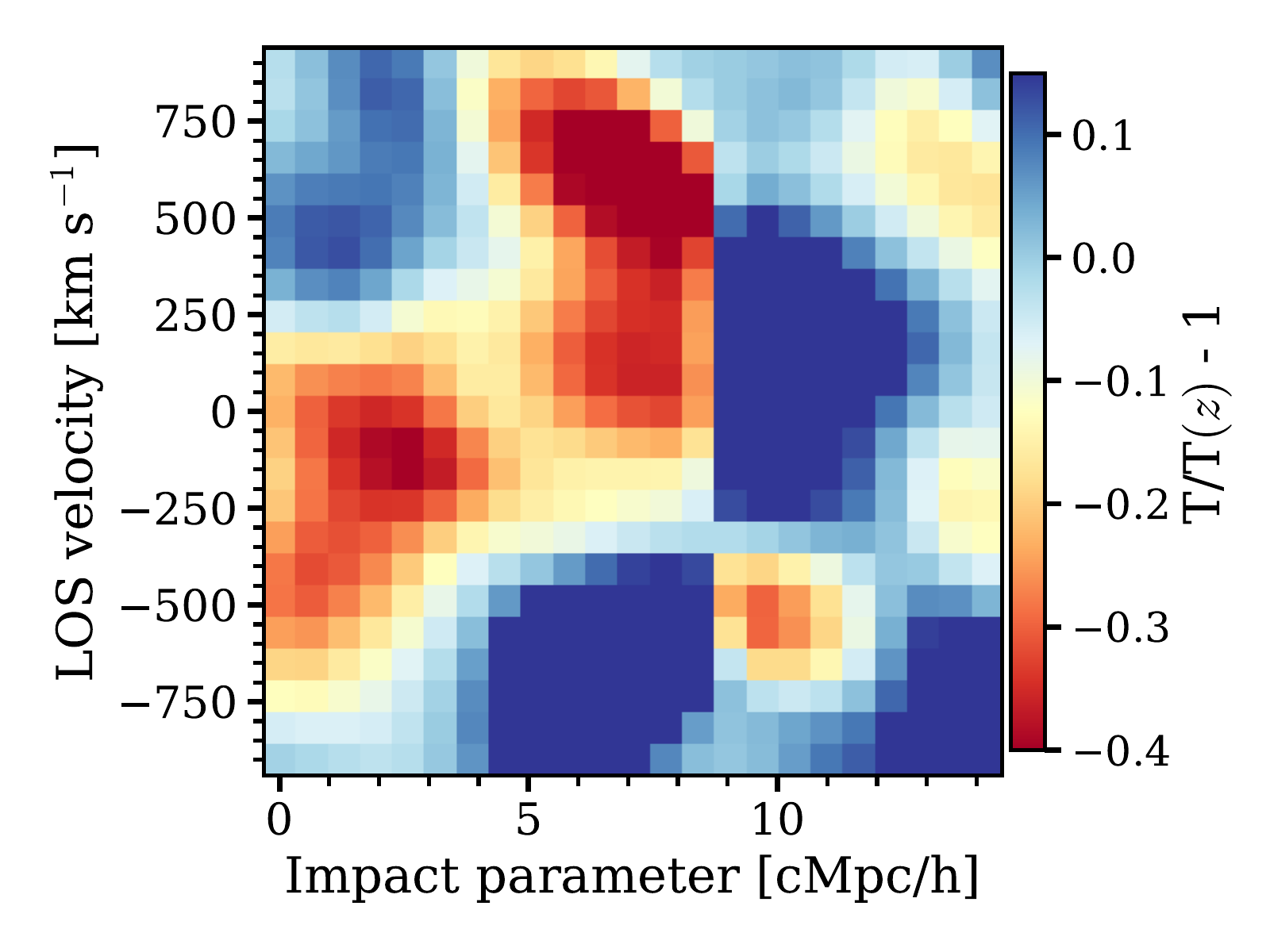}
       & \includegraphics[width=7cm]{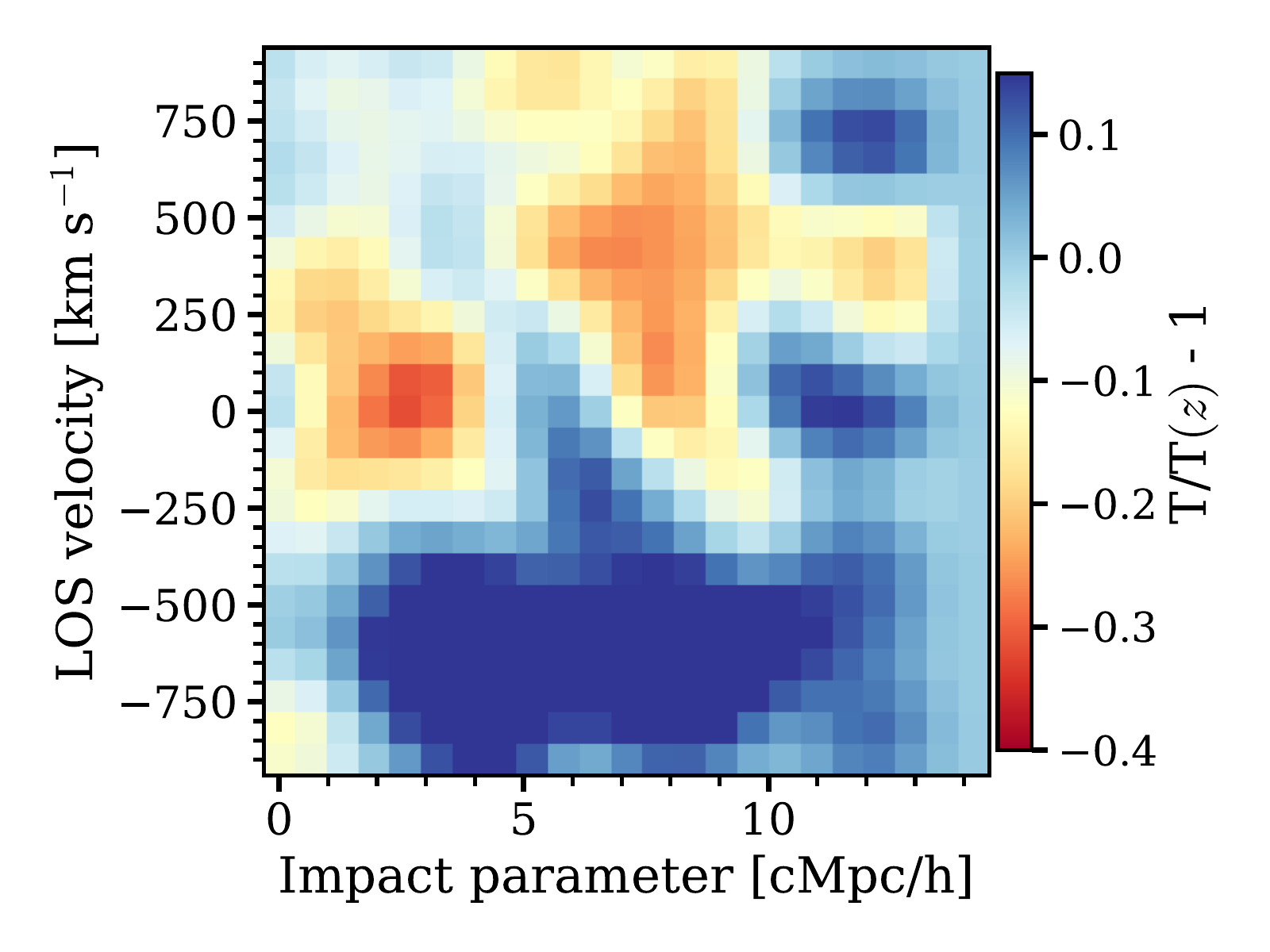} \\
        \includegraphics[width=7cm]{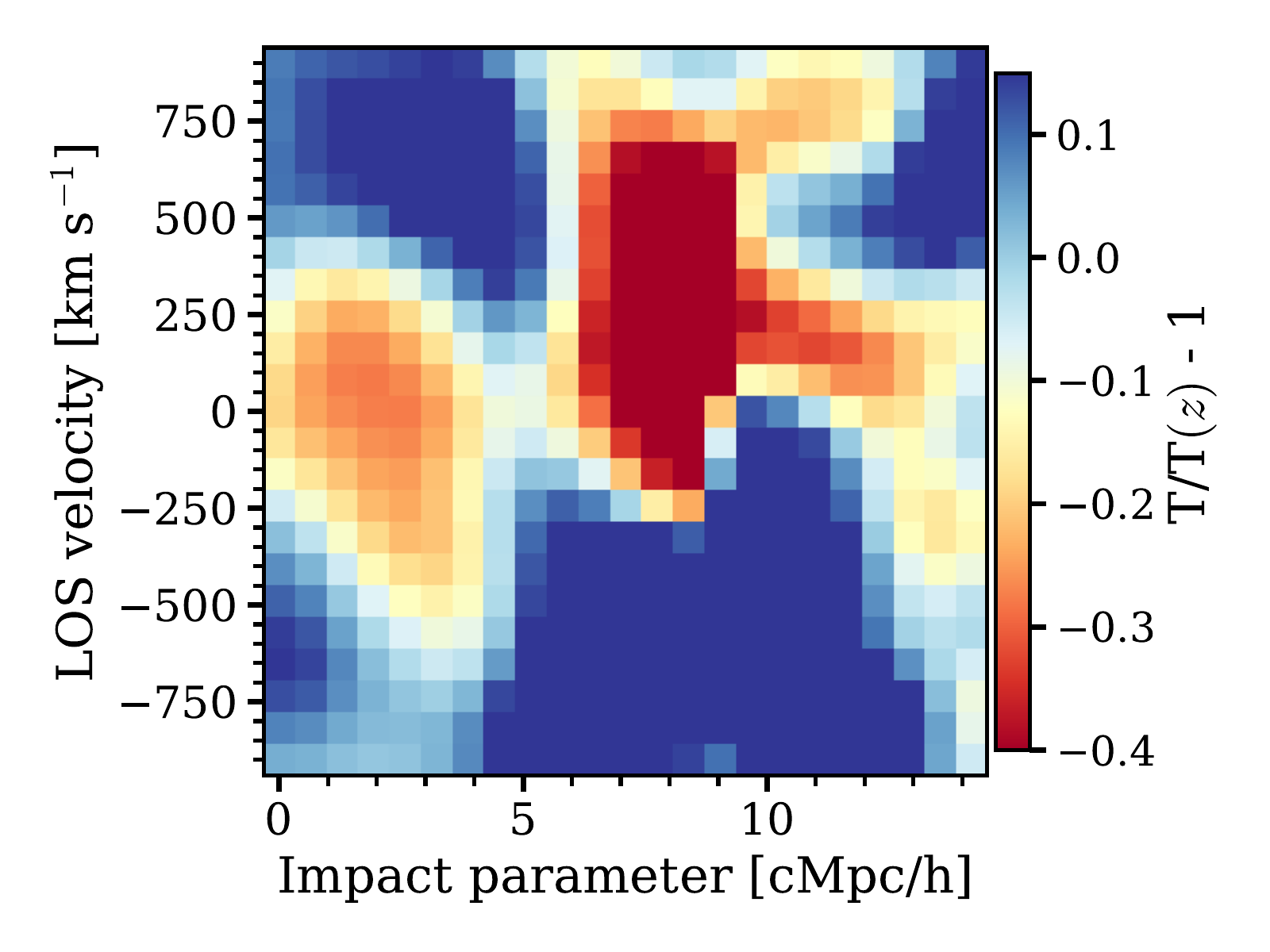}
       & \includegraphics[width=7cm]{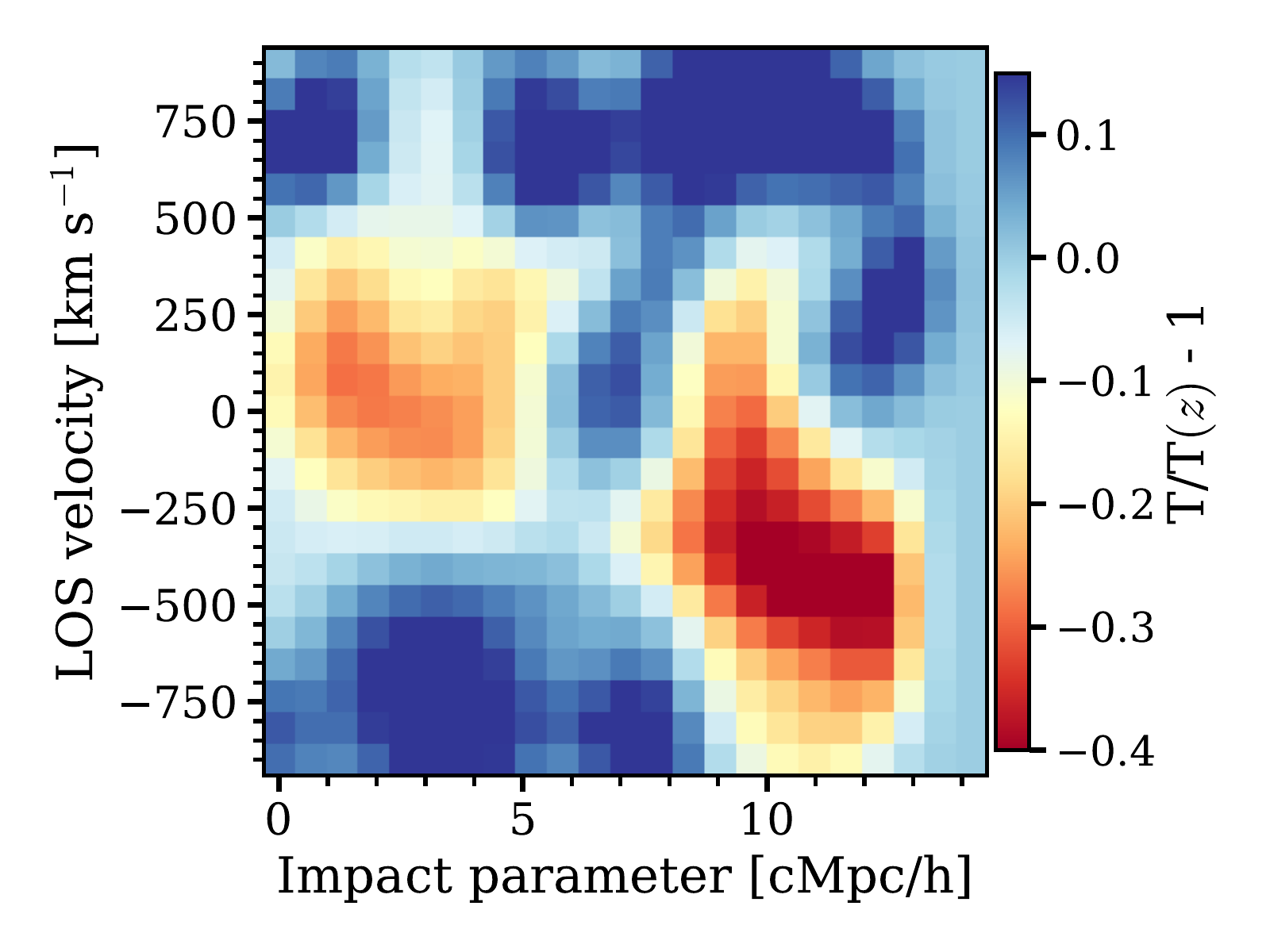} \\
    \end{tabular}
    \caption{The excess transmission as a function of the impact parameter and the LOS velocity (as in Fig. $\ref{fig:2Dplot}$), where the foreground sample is split in four quadrants. }
    \label{fig:JACKS}
\end{figure*}

\begin{figure}
    \centering
    \includegraphics[width=8cm]{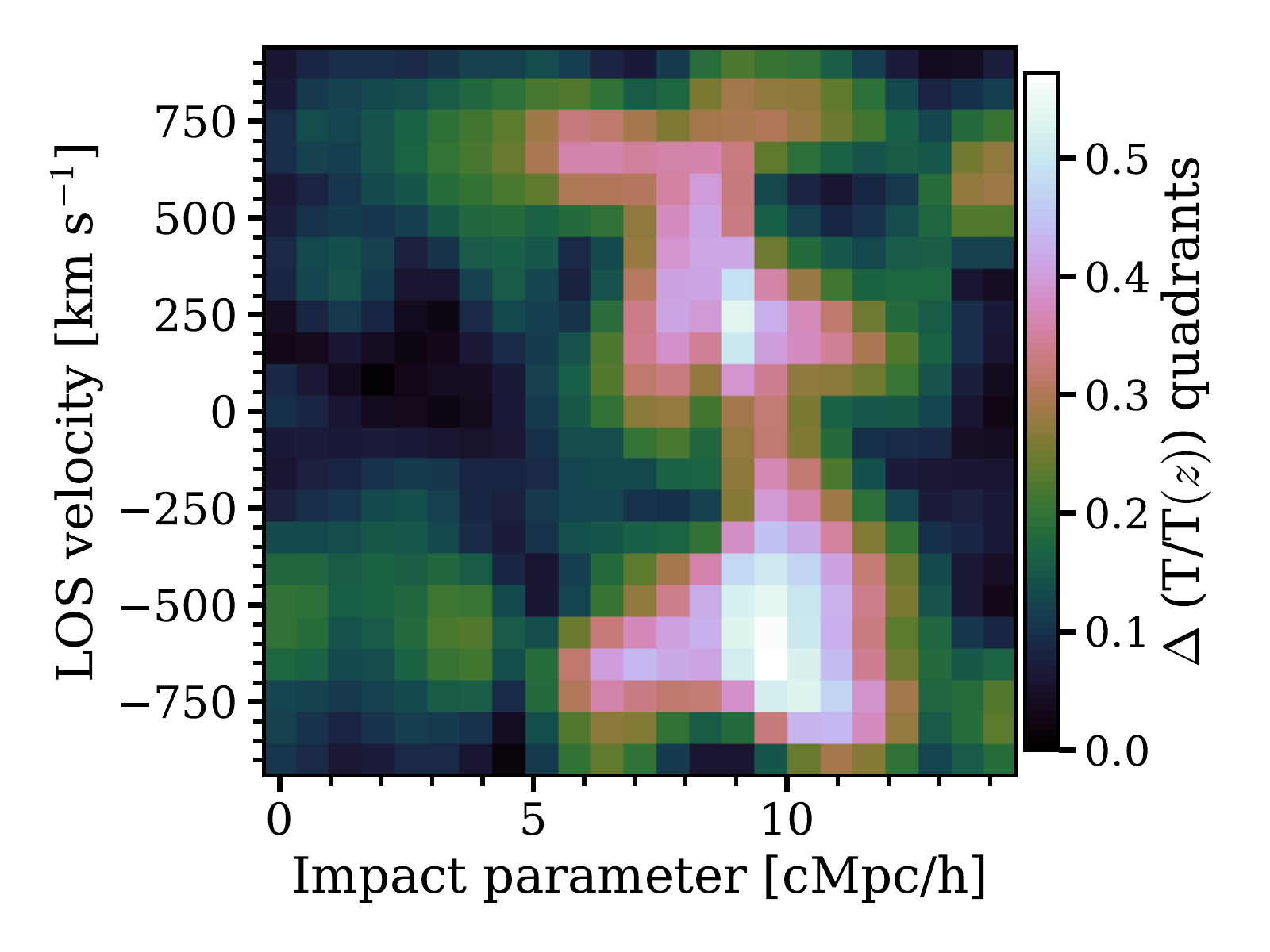}
    \caption{The maximum absolute difference among the excess transmission in the four quadrants as a function of the impact parameter and the LOS velocity (as Fig. $\ref{fig:2Dplot}$). The inhomogeneity of the survey design impacts the results most at the impact parameters $\sim10$ cMpc/h$_{70}$.}
    \label{fig:std_2D}
\end{figure}

As shown in Fig. $\ref{fig:RADEC}$, our survey design is rather inhomogeneous in terms of sensitivity and coverage. This impacts the distribution of impact parameters of the foreground galaxies (and their luminosities). In this section, we investigate the robustness of the measured cross-correlation signal to effects related to this inhomogeneity. 
To test the gaps and holes in the survey design, we have split our foreground sample in four quadrants whose division lines center on ID53 and are orthogonal and parallel to the major extent of the larger MUSE-Wide region (see Fig. $\ref{fig:RADEC}$). We then redo the cross-correlation analysis in each of these quadrants individually. In Fig. $\ref{fig:CC_jacks}$ we show the excess HI transmission as a function of 3D distance for each of these four quadrants. There are no significant differences between the quadrants within 10 cMpc/h$_{70}$. In Figures $\ref{fig:JACKS}$, we separate the transverse and the line of sight distances. The excess absorption is consistently detected within $\sim3$ cMpc/h$_{70}$ and $\sim200$ km s$^{-1}$ along the line of  sight. In Fig.  $\ref{fig:std_2D}$, we show the largest absolute difference in the excess transmission among the four quadrants at each distance. This illustrates that the scatter due to the survey design is largest at impact parameters $\sim10$ cMpc/h$_{70}$, precisely the range where the survey layout is most patchy. Nevertheless, the main result of excess absorption within $\lesssim3$ cMpc/h$_{70}$ from galaxies is robust.

\begin{figure}
    \centering
    \includegraphics[width=8cm]{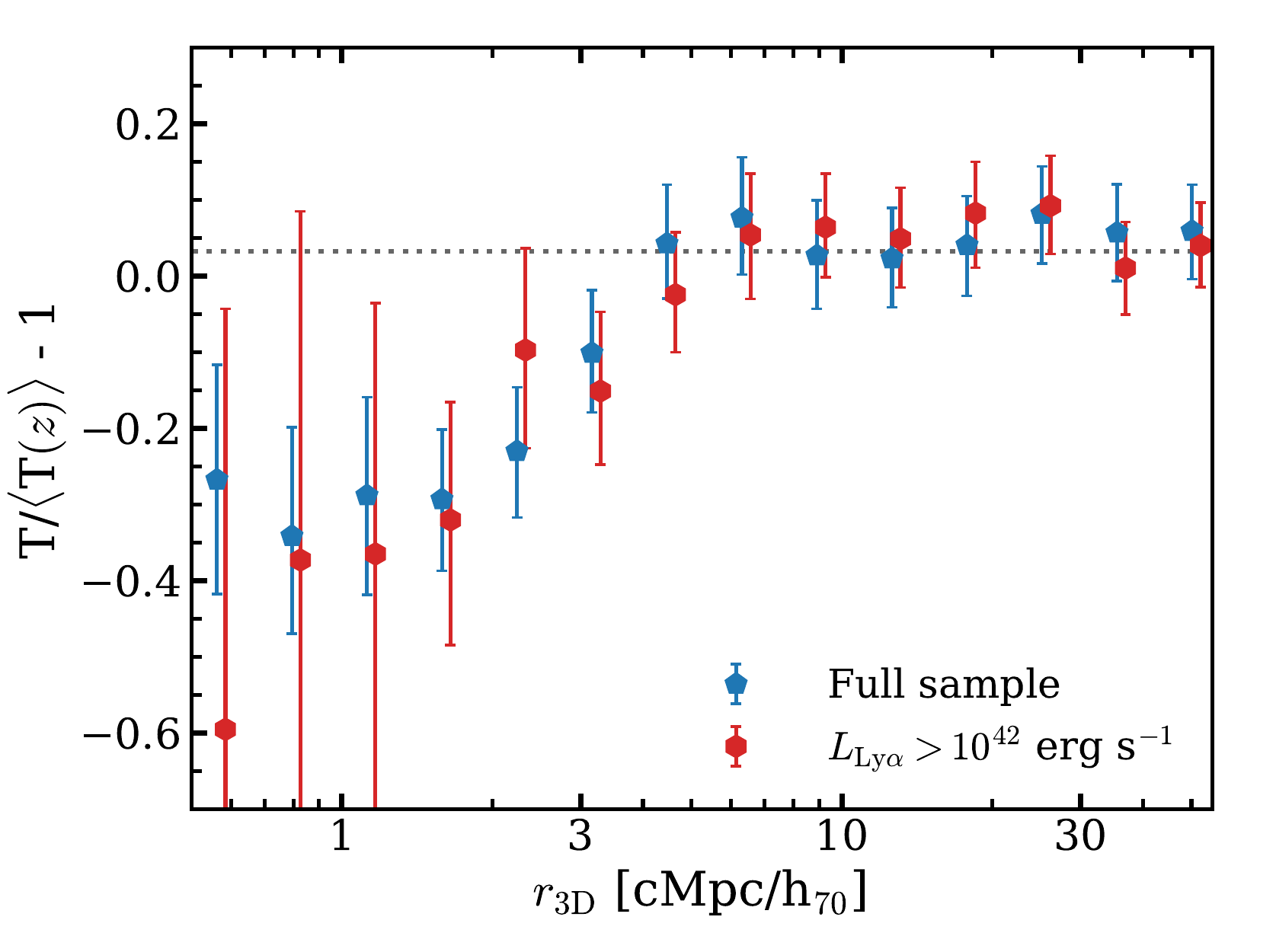}
    \caption{The excess HI transmission as a function of 3D distance from the foreground galaxies (as in Fig. $\ref{fig:3D}$) when the full sample is considered (blue points with error bars) and when adopting a uniform luminosity cut L$_{\rm Ly\alpha} > 10^{42}$ erg s$^{-1}$ (red points with error bars). }
    \label{fig:MW}
\end{figure}

We also test the impact of the varying sensitivity of the foreground sample (see bottom panel in Fig. $\ref{fig:stats}$). In Fig. $\ref{fig:MW}$, we show the excess transmission as a function of 3D distance for the full sample (the result shown in Fig. $\ref{fig:3D}$), and limiting our sample to Ly$\alpha$ luminosities $>10^{42}$ erg s$^{-1}$ which are detectable over the full survey field. While in the second case the uncertainty increases substantially at the smallest impact parameters due to the lower number density of foreground galaxies, the general trend is fully consistent with the one derived from the full sample.

\section{Other possible background sources} \label{app:depths}

\begin{figure*}
    \centering
    \includegraphics[width=16cm]{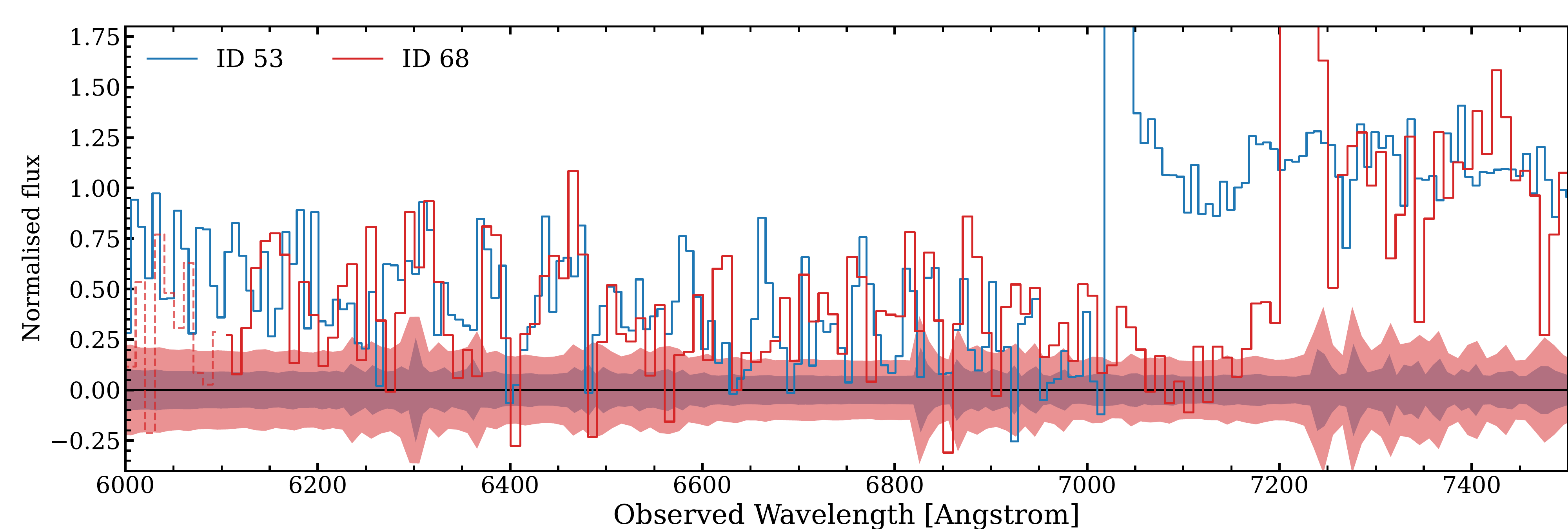}
    \caption{The Ly$\alpha$ forest transmission and Ly$\alpha$ emission spectrum of the galaxy ID53 (main paper, blue) compared to galaxy ID68, which is a slightly fainter galaxy at $z=4.9$ within the MXDF coverage (red). The line-style is changed to dashed for regions that are impacted by Ly$\beta$ forest absorption. Spectra are binned to facilitate the visual comparisons. The shaded regions show the noise levels. Here, for simplicity, spectra are normalised to the median flux at their respective rest-frame wavelength 1250-1450 {\AA}. } 
    \label{fig:ID68}
\end{figure*}

As mentioned in \S $\ref{sec:future}$, there are several other galaxies with a comparable continuum UV magnitude to our main target at redshifts $z\sim5$ in the MUSE coverage. While the MUSE Wide region has a particularly bright galaxy at $z=4.84$ with m$_{1500}=23.3$, this galaxy has a relatively red UV continuum slope (the F105W-F125W colour is +0.5) that indicates significant dust attenuation in the ISM of the galaxy. As a consequence of this and the mere 1hr exposure time of the MUSE Wide survey, the intrinsic flux of the galaxy in the Lyman-$\alpha$ forest region is too faint for this galaxy to be useful for Lyman-$\alpha$ absorption studies.

In the t$_{\rm exp}$ = 10 hr MOSAIC regions, there are two galaxies (IDs 1185 and 1264) at $z=4.5$ and $z=4.8$, respectively, which are 0.1 and 0.4 magnitudes more luminous than ID53. However, one of these galaxies is strongly contaminated by a foreground galaxy, while the other has a red SED with various absorption features, similar to the MUSE Wide galaxy.

Within the deep MXDF region, there is a galaxy (ID68, at $z=4.94$; \citealt{Bacon22}) which is 0.4 magnitude fainter than ID53. It is separated to ID53 by 22$''$ on sky ($\approx0.8$ cMpc/h$_{70}$). In Fig. $\ref{fig:ID68}$, we compare the Ly$\alpha$ forest spectrum of this galaxy to the spectrum of ID53, both simply normalised to their rest-frame median flux at $\lambda=1250-1450$ {\AA}. It appears that the Lyman-$\alpha$ transmission in the two spectra correlates quite well at observed wavelengths $\lambda_{\rm obs} = 6200 - 6600$ {\AA} (which corresponds to $z=4.1-4.4$). The correlation between the absorption spectra weaken at higher redshifts, which could be real, but also due to the relatively higher noise in the spectrum (caused by skylines). With a more detailed analysis (in particular of the intrinsic SED model), our tomographic experiment could independently be repeated, albeit with somewhat higher uncertainties. One could use the combination of the two sight-lines to attempt to constrain the coherence of individual IGM structures over the $\sim1$ cMpc/h$_{70}$ scale, which we leave for future work.

These comparisons highlight that, in addition to being UV bright, the best galaxies for comparable tomography studies are those with a young age and a low dust attenuation as they have a relatively featureless (blue) continuum in the Lyman-$\alpha$ forest region.


\bsp	
\label{lastpage}
\end{document}